\documentclass[twocolumn,times,twocolappendix]{aastex63}

\usepackage{CJKutf8}
\usepackage{amsmath}
\usepackage[T1]{fontenc}
\usepackage{textcomp}

\submitjournal{ApJ}

\shorttitle{On the crystallinity of silicate dust in evolving PPDs due to MDWs}
\shortauthors{Arakawa et al.}
\graphicspath{{./}{figures/}}
\definecolor{dark-green}{rgb}{0.0, 0.5, 0.0}

\begin{document}

\title{
On the crystallinity of silicate dust in evolving protoplanetary disks due to magnetically driven disk winds}

\correspondingauthor{Sota Arakawa}
\email{sota.arakawa@nao.ac.jp}

\author[0000-0003-0947-9962]{Sota Arakawa}
\affiliation{Division of Science, National Astronomical Observatory of Japan \\
2-21-1 Osawa, Mitaka, Tokyo 181-8588, Japan.}

\author[0000-0002-2383-1216]{Yuji Matsumoto}
\affiliation{Center for Computational Astrophysics, National Astronomical Observatory of Japan \\
2-21-1 Osawa, Mitaka, Tokyo 181-8588, Japan.}

\author[0000-0002-6172-9124]{Mitsuhiko Honda}
\affiliation{Faculty of Biosphere--Geosphere Science, Okayama University of Science \\
1-1 Ridai-chou, Okayama 700-0005, Japan.}

%% Note that the \and command from previous versions of AASTeX is now
%% depreciated in this version as it is no longer necessary. AASTeX 
%% automatically takes care of all commas and "and"s between authors names.

%% AASTeX 6.3 has the new \collaboration and \nocollaboration commands to
%% provide the collaboration status of a group of authors. These commands 
%% can be used either before or after the list of corresponding authors. The
%% argument for \collaboration is the collaboration identifier. Authors are
%% encouraged to surround collaboration identifiers with ()s. The 
%% \nocollaboration command takes no argument and exists to indicate that
%% the nearby authors are not part of surrounding collaborations.

%% Mark off the abstract in the ``abstract'' environment.

\begin{abstract}
We present a novel mechanism for the outward transport of crystalline dust particles: the outward radial drift of pebbles.
The dust ring structure is frequently observed in protoplanetary disks.
One of the plausible mechanisms of the formation of dust rings is the accumulation of pebbles around the pressure maximum, which is formed by the mass loss due to magnetically driven disk winds.
In evolving protoplanetary disks due to magnetically driven disk winds, dust particles can migrate outwardly from the crystallization front to the pressure maximum by radial drift.
We found that the outward radial drift process can transport crystalline dust particles efficiently when the radial drift timescale is shorter than the advection timescale.
Our model predicts that the crystallinity of silicate dust particles could be as high as 100\% inside the dust ring position.
\end{abstract}

%% Keywords should appear after the \end{abstract} command. 
%% See the online documentation for the full list of available subject
%% keywords and the rules for their use.
%%\keywords{editorials, notices --- miscellaneous --- catalogs --- surveys}

%% From the front matter, we move on to the body of the paper.
%% Sections are demarcated by \section and \subsection, respectively.
%% Observe the use of the LaTeX \label
%% command after the \subsection to give a symbolic KEY to the
%% subsection for cross-referencing in a \ref command.
%% You can use LaTeX's \ref and \label commands to keep track of
%% cross-references to sections, equations, tables, and figures.
%% That way, if you change the order of any elements, LaTeX will
%% automatically renumber them.
%%
%% We recommend that authors also use the natbib \citep
%% and \citet commands to identify citations.  The citations are
%% tied to the reference list via symbolic KEYs. The KEY corresponds
%% to the KEY in the \bibitem in the reference list below. 

%@arxiver{fig_schematic.pdf,fig_sigmad_alpha1e-4_Sc1e0_vf3e2_Cw1e-5.eps,fig_C_alpha1e-4_Sc1e0_vf3e2_Cw1e-5.eps}

\section{Introduction}
\label{sec:intro}

Protoplanetary disks are the birth places of planetary systems.
Therefore, disk evolution is of great importance to understand how planets formed \citep[e.g.,][]{2014prpl.conf..339T}.
As planetesimal formation via accumulation of dust particles is the first step of planet formation \citep[e.g.,][]{2014prpl.conf..547J}, the spatial distribution and migration of dust particles in evolving protoplanetary disks have been studied extensively so far.

Several pieces of evidence suggest that silicate dust undergoes significant thermal processing in protoplanetary disks.
In the interstellar medium, silicate dust is thought to be completely amorphous, as suggested by broad and smooth absorption features \citep[e.g.,][]{2004ApJ...609..826K}.
In contrast, crystalline silicate features are often found toward disks around Herbig Ae/Be stars \citep[e.g.,][]{1995ApJ...438..250H} and T Tauri stars \citep[e.g.,][]{2003ApJ...585L..59H,2006ApJ...646.1024H}.
Forsterite (${\rm Mg}_{2}{\rm Si}{\rm O}_{4}$) is the most abundant silicate mineral in disks \citep[e.g.,][]{2010ApJ...721..431J,2015A&A...574A.140M}.
Laboratory experiments suggest that the crystallization temperature of forsterite is approximately 600--1000 K \citep[e.g.,][]{2000ApJ...535..247H,2018ECS.....2..778Y}.
However, in some disks around young stars, crystalline forsterite has been observed much farther away from the ``crystallization front'', where the disk temperature is equal to the crystallization temperature \citep[e.g.,][]{2010ApJ...721..431J,2012Natur.490...74D,2013A&A...553A...5S}.
In addition, comets and interplanetary dust particles in our solar system contain crystalline silicate \citep[e.g.,][]{2004ApJ...601..577H,2007P&SS...55.1044O,2009M&PS...44.1675O}.
These facts strongly suggest that the crystallization and outward transport processes occurred both in extrasolar protoplanetary disks and in the early solar nebula.

Several mechanisms have been suggested as driving the outward transport of crystalline dust particles in protoplanetary disks, including turbulent diffusion \citep[e.g.,][]{2001A&A...378..192G,2010Icar..208..455C}, large-scale circulations associated with mass and angular momentum transfer \citep[e.g.,][]{2004A&A...415.1177K,2007Sci...318..613C,2009Icar..200..655C}, spiral arms in gravitationally unstable massive disks \citep[e.g.,][]{2008E&PSL.268..102B}, photophoresis \citep[e.g.,][]{2007A&A...466L...9M}, and radiation pressure \citep[e.g.,][]{2009Natur.459..227V,2015ApJ...799..119T}.
\citet{2006ApJ...640L..67D} proposed that the majority of crystalline dust particles are formed in the very early phase of disk formation via the collapse of molecular cloud cores \citep[see also][]{2012M&PS...47...99Y}.
In addition, parts of crystalline dust particles might be formed in-situ in the outer region of protoplanetary disks by exothermic chemical reactions of reactive molecules \citep{2010ApJ...717..586T} and/or by shock waves \citep[e.g.,][]{2002ApJ...565L.109H,2010ApJ...719..642M}.

The physical process of outward transport by turbulence is described by the diffusion equation.
The key parameter for turbulent mixing in steady state disks is the Schmidt number, which is the ratio of the kinetic viscosity and diffusion coefficient \citep[e.g.,][]{1988MNRAS.235..365C}.
\citet{2007A&A...471..833P} reviewed how the Schmidt number affects the crystallinity of disks.
They found that the radial distribution of the crystallinity in steady state disks is given by a power-law distribution, and the exponent depends on the Schmidt number.
A small value of the Schmidt number lower than one is required for efficient outward diffusion in standard accretion disks \citep[e.g.,][]{1988MNRAS.235..365C,2007A&A...471..833P,2010ApJ...719.1633H}.

The radial drift of dust particles due to gas drag is another important process to understand the radial transport of dust particles.
In classical views of accretion disks \citep[e.g.,][]{1974MNRAS.168..603L,1998ApJ...495..385H}, the pressure gradient at the midplane is negative throughout the disk, and dust particles migrate inward due to gas drag \citep[e.g.,][]{1976PThPh..56.1756A}.
This inward drift offsets the outward transport of dust particles \citep[e.g.,][]{2010Icar..208..455C}.
In addition, mm- to cm-sized large dust particles spiral into the central star within 1~Myr unless a local pressure maximum prevents the dust particles from inward migration \citep[e.g.,][]{2018ApJS..238...11D,2021ApJ...908...64F}.
Therefore, it is difficult for standard accretion disks to transport dust particles formed at high temperatures outwardly and to maintain high crystallinity for a long time in the outer region.
We briefly review the effect of radial drift on the radial distribution of the crystallinity in Section \ref{sec:no-wind}.

Recent astronomical observations have revealed varieties of structures of protoplanetary disks \citep[e.g.,][]{2013PASJ...65L..14F,2015ApJ...808L...3A,2016ApJ...829L..35T,2017ApJ...837..132V,2018ApJ...869L..41A}.
The observed disk structures provide us plenty of clues to reveal how planets formed \citep[see][and references therein]{2020ARA&A..58..483A}.
In particular, the dust ring structure is observed in a large number of disks. 
To date, several mechanisms are proposed for the origin: planets \citep[e.g.,][]{2015ApJ...809...93D,2018ApJ...868...48K}; dust growth \citep[e.g.,][]{2014A&A...572A.107L,2021ApJ...907...80O}; condensation fronts \citep[e.g.,][]{2016ApJ...821...82O,2017ApJ...845...68P}; photoevaporative flows \citep[e.g.,][]{2017RSOS....470114E}; disk instabilities due to dust--gas friction and self-gravity (or turbulent gas viscosity) \citep[e.g.,][]{2014ApJ...794...55T,2019ApJ...881...53T}; magnetically driven disk winds \citep[e.g,][]{2018ApJ...865..102T}.

In this study, we focus on the crystallinity of silicate dust particles in ring structures formed by the magnetically driven disk winds.
\citet{2009ApJ...691L..49S} found that magneto-hydrodynamic turbulence in protoplanetary disks drives disk winds.
\citet{2010ApJ...718.1289S} revealed that the mass loss timescale of the magnetically driven disk winds is proportional to the local Keplerian rotation period, and disk winds disperse the gas component of disks from the inner region.
In other words, magnetically driven disk winds potentially create a maximum of gas pressure around 1--10~au from the central stars \citep[e.g.,][]{2016A&A...596A..74S}, and dust particles which are dynamically decoupled with gas (referred to as ``pebbles'') are accumulated around the pressure maximum \citep[e.g.,][]{2003ApJ...583..996H}.
This is the formation mechanism of dust ring in evolving disk due to disk winds proposed by  \citet{2018ApJ...865..102T}.

In this study, we propose a novel mechanism for outward transport of crystalline dust particles.
In evolving protoplanetary disks due to magnetically driven disk winds, dust particles can migrate outwardly by radial drift.
We found that the {\it outward radial drift} process can transport crystalline dust particles efficiently when the radial drift overcomes the advective flow (see Figure \ref{fig:sch}).
Our model predicts that the crystallinity of silicate dust particles could be as high as 100\% inside the dust ring position, and this is totally different from the prediction for accretion disks without disk winds \citep[e.g.,][]{2007A&A...471..833P}.

\section{Models}
\label{sec:model}

In Section \ref{sec:model}, we briefly introduce the equations used to compute the evolution of protoplanetary disks.
We calculate the temporal evolution of the surface densities of gas and dust using vertically integrated disk models, and we also obtain the radial distribution of the crystallinity of dust particles.
The basic equations for the evolution of gas and dust disks are described in Sections \ref{sec:gasdisk} and \ref{sec:dustdisk}, respectively.

\subsection{Evolution of gas disk}
\label{sec:gasdisk}

We set the initial distribution of the gas surface density, $\Sigma_{{\rm gas}, 0}$, as a self-similar profile, which is described as follows \citep{1974MNRAS.168..603L}:
{\small
\begin{equation}
\Sigma_{{\rm gas}, 0} = \frac{{\left( 2 - \gamma \right)} M_{\rm disk}}{2 \pi {r_{0}}^{2}} {\left( \frac{r}{r_{0}} \right)}^{- \gamma} \exp{\left[ {\left( - \frac{r}{r_{0}} \right)}^{2 - \gamma} \right]},
\label{eq:Sigmagini}
\end{equation}
}where $M_{\rm disk} = 0.01 M_{\star}$ is the total mass of the gas disk, $r_{0} = 100\ {\rm au}$ is the initial disk radius, and $\gamma = 1$ is the exponent for the gas surface density profile.
Here $r$ denotes the distance from the central star.
As an example, we take a Herbig Ae/Be star with the mass of $M_{\star} = 2.5 M_{\odot}$ ($M_{\odot}$ is the solar mass) as assumed in \citet{2007A&A...471..833P}.
For simplicity, we assume that the (midplane) temperature of the disk are given as follows:
\begin{equation}
T = T_{\rm 1} {\left( \frac{r}{1\ {\rm au}} \right)}^{- q},
\end{equation}
where $T_{\rm 1} = 800\ {\rm K}$ is the temperature at $r = 1\ {\rm au}$, and $q = 1/2$ is the exponent for the temperature structure.
We also set the location of the crystallization front at $r_{\rm c} = 1\ {\rm au}$, and all dust is crystalline for $r \leq r_{\rm c}$ \citep{2007A&A...471..833P}.

The basic equation of the evolution of gas surface density of accretion disks with magnetically driven disk winds is
\begin{equation}
\frac{\partial \Sigma_{\rm gas}}{\partial t}  = \frac{1}{2 \pi r} \frac{\partial \dot{M}_{\rm gas}}{\partial r} + \dot{\Sigma}_{\rm wind},
\label{eq:Sigmag}
\end{equation}
where 
\begin{equation}
\dot{M}_{\rm gas} = 6 \pi r^{1/2} \frac{\partial {\left( r^{1/2} \Sigma_{\rm gas} \nu \right)}}{\partial r},
\end{equation}
is the (vertically integrated) mass flux at every location $r$, and $\dot{\Sigma}_{\rm wind}$ is the mass loss rate due to the disk wind.
Then, the advection velocity, $v_{\rm adv}$, is
\begin{equation}
v_{\rm adv} = \frac{\dot{M}_{\rm gas}}{2 \pi r \Sigma_{\rm gas}}.
\end{equation}
The advection velocity is positive when the gas flows inwardly.

The mass flux is proportional to the kinematic viscosity,
\begin{equation}
\nu = \alpha_{\rm acc} c_{\rm s} h_{\rm g},
\end{equation}
where $\alpha_{\rm acc}$ is the angular momentum transport efficiency parameter called alpha parameter \citep{1973A&A....24..337S}, $c_{\rm s}$ is the sound speed, and $h_{\rm g}$ is the gas scale height.
The gas scale height, $h_{\rm g}$, and the midplane gas density, $\rho_{\rm g}$, are given by
\begin{eqnarray}
h_{\rm g} & = & \frac{c_{\rm s}}{\Omega_{\rm K}}, \\
\rho_{\rm g} & = & \frac{\Sigma_{\rm gas}}{\sqrt{2 \pi} h_{\rm g}},
\end{eqnarray}
where $\Omega_{\rm K} = \sqrt{{G M_{\star}}/{r^{3}}}$ is the Keplerian frequency, and $G$ is the gravitational constant.

\citet{2010ApJ...718.1289S} investigated the mass loss rate due to the magnetically driven disk wind.
Based on their three-dimensional local magnetohydrodynamic simulations, the mass loss rate due to the disk wind is given by
\begin{equation}
\dot{\Sigma}_{\rm wind} = - C_{\rm w} \Sigma_{\rm gas} \Omega_{\rm K},
\end{equation}
where $C_{\rm w}$ is the efficiency parameter \citep[see also][]{2018ApJ...865..102T}.
We set the typical value of $C_{\rm w} = 10^{-5}$ in Section \ref{sec:wind}.

\citet{2016ApJ...821....3M} found that not only gas but also small dust particles can be blown out by the disk wind. 
However, we do not consider this effect.
The dust blown-out process works when their Stokes number satisfies ${\rm St} < C_{\rm w} / 1.8$ \citep{2021ApJ...909...75T}.
As we set $C_{\rm w} = 10^{-5}$, pebbles which can drift due to gas drag would not be blown out by the disk wind.

\subsection{Motion of dust particles}
\label{sec:dustdisk}

We set the initial distribution of the dust surface density, $\Sigma_{{\rm dust}, 0}$, as follows:
\begin{equation}
\Sigma_{{\rm dust}, 0} = 0.01 \Sigma_{{\rm gas}, 0}.
\end{equation}
In this study, we consider two types of dust particles: crystalline and amorphous particles.
We define the crystallinity, $C$, as the fraction of the crystalline dust particles:
\begin{equation}
C \equiv \frac{\Sigma_{\rm c}}{\Sigma_{\rm dust}},
\end{equation}
where $\Sigma_{\rm c}$ is the surface density of the crystalline dust particles, and the dust  surface density is the sum of the surface densities of crystalline and amorphous particles:
$\Sigma_{\rm dust} = \Sigma_{\rm c} + \Sigma_{\rm a}$.

We compute the temporal evolution of the surface densities of crystalline and amorphous dust particles.
We consider three physical processes: advection in the mean gas flow, diffusion due to concentration gradient, and the radial drift of dust particles relative to the gas \citep[see][and references therein]{2021ApJ...908...64F}.
In addition, we also take into account the effect of conversion of dust to planetesimals via the streaming instability (see Section \ref{sec:SI}).

The surface densities of crystalline and amorphous dust particles evolve according to
\begin{equation}
\frac{\partial \Sigma_{i}}{\partial t}  = \frac{1}{2 \pi r} \frac{\partial \dot{M}_{i}}{\partial r} + \dot{\Sigma}_{{\rm plts}, i},
\end{equation}
where the subscript $i$ denotes the crystalline ($i = {\rm c}$) or amorphous ($i = {\rm a}$) dust particles.
The mass flux of dust particles, $\dot{M}_{i}$, is given by the sum of the three physical processes:
\begin{equation}
\dot{M}_{i} = \dot{M}_{i, {\rm adv}} + \dot{M}_{i, {\rm diff}} + \dot{M}_{i, {\rm drift}},
\end{equation}
and $\dot{\Sigma}_{{\rm plts}, i}$ is the conversion rate of dust to planetesimals.

\subsubsection{Advection, diffusion, and radial drift}
\label{sec:motion}

\citet{2017ApJ...840...86D} re-derived the equations for radial transport of dust particles.
The advection term is given by
\begin{equation}
\dot{M}_{i, {\rm adv}} = 2 \pi r \Sigma_{i} v_{\rm adv},
\end{equation}
and the diffusion term is
\begin{equation}
\dot{M}_{i, {\rm diff}} = 2 \pi r \Sigma_{i} v_{i, {\rm diff}},
\end{equation}
where the diffusion velocity, $v_{i, {\rm diff}}$, is given by
\begin{equation}
v_{i, {\rm diff}} = D {\left( \frac{\Sigma_{i}}{\Sigma_{\rm gas}} \right)}^{-1} {\frac{\partial}{\partial r} {\left( \frac{\Sigma_{i}}{\Sigma_{\rm gas}} \right)}}.
\end{equation}
Here $D$ is the diffusion coefficient of the dust particles, which is given by
\begin{equation}
D = \frac{\nu}{{\rm Sc} {\left( 1 + {\rm St}^{2} \right)}},
\label{eq:D}
\end{equation}
where ${\rm Sc}$ is the Schmidt number and ${\rm St}$ is the Stokes number of the dust particles (see Section \ref{sec:Vfrag}).
The diffusion velocity is positive when the direction of the flow is inward.
We note that the Stokes number is sufficiently small (${\rm St} \ll 1$) in our simulations, and the diffusion coefficient is approximately given by $D \simeq \nu / {\rm Sc}$.
The radial drift term is given by
\begin{equation}
\dot{M}_{i, {\rm drift}} = 2 \pi r \Sigma_{i} v_{\rm drift},
\end{equation}
and the drift velocity, $v_{\rm drift}$, is given by the following equation:
\begin{equation}
v_{\rm drift} = \frac{{\rm St}}{1 + {\rm St}^{2}} {\left( \eta r \Omega_{\rm K} - {\rm St} v_{\rm adv} \right)}.
\end{equation}
Here $\eta$ is the normalized pressure gradient, which is given by
\begin{equation}
\eta = - \frac{1}{r {\Omega_{\rm K}}^{2}} \frac{1}{\rho_{\rm g}} \frac{\partial P}{\partial r},
\end{equation}
and $P = \rho_{\rm g} {c_{\rm s}}^{2}$ is the gas pressure at the midplane.
We note that $v_{\rm drift}$ is not the radial velocity of dust particles toward the central star but the radial drift velocity relative to the gas \citep[see][]{2017ApJ...840...86D}.
The radial velocity of dust particles toward the central star is the sum of $v_{\rm adv}$ and $v_{\rm drift}$.

\subsubsection{Stokes number of dust particles}
\label{sec:Vfrag}

The Stokes number is the key parameter for the radial drift of dust particles and controls the velocities of dust particles.
Assuming that fragmentation limits dust growth\footnote{
We note that radial drift toward the central star may limit dust growth when the Stokes number exceeds $\sim 10^{-1}$, even if we do not consider fragmentation \citep[e.g.,][]{2012ApJ...752..106O,2019ApJ...878..132O}.
}, the Stokes number of dust particles is given by the equilibrium between the mutual collision velocities and the fragmentation velocity:
\begin{equation}
{\Delta v} = v_{\rm frag},
\label{eq:vfrag}
\end{equation}
where ${\Delta v}$ is the mutual collision velocity, which depends on ${\rm St}$, and $v_{\rm frag}$ is the threshold velocity for collisional fragmentation/growth.
The mutual collision velocity is given by
\begin{equation}
{\left( \Delta v \right)}^{2} = {\left( \Delta v_{\rm r} \right)}^{2} + {\left( \Delta v_{\rm t} \right)}^{2},
\label{eq:DeltaV}
\end{equation}
where $\Delta v_{\rm r}$ and $\Delta v_{\rm t}$ are the contributions from radial drift \citep[e.g.,][]{1976PThPh..56.1756A} and gas turbulence \citep[e.g.,][]{2007A&A...466..413O}, respectively.
For the case of $10^{-4} \lesssim {\rm St} \ll 1$, \citet{2016ApJ...821...82O} found that the radial drift term is given by 
\begin{equation}
{\Delta v_{\rm r}} \simeq 0.5 {\rm St} \eta r \Omega_{\rm K},
\end{equation}
and the gas turbulence term is given by
\begin{equation}
{\Delta v_{\rm t}} \simeq \sqrt{2.3 \alpha_{\rm turb} {\rm St}} c_{\rm s},
\label{eq:DeltaVt}
\end{equation}
where $\alpha_{\rm turb}$ is the dimensionless parameter for the strength of turbulence.
Then, we can calculate the Stokes number from Equation (\ref{eq:DeltaV}), which is the quadratic equation for ${\rm St}$.

The strength of turbulence should be associated with the strength of mass diffusion.
Therefore, the two alpha parameters, $\alpha_{\rm turb}$ and $\alpha_{\rm acc}$, might be related with the Schmidt number, which is the ratio of the kinematic viscosity to the mass diffusion coefficient.
We simply assume the following equation:
\begin{equation}
\alpha_{\rm turb} = \frac{\alpha_{\rm acc}}{{\rm Sc}},
\end{equation}
although we set ${\rm Sc} = 1$ in the main part of this study otherwise noted.
Hence, $\alpha_{\rm turb} = \alpha_{\rm acc}$ is assumed.

We assume that vertical settling of dust particles balances with turbulent diffusion.
The dust scale height, $h_{\rm d}$, and the midplane dust density, $\rho_{\rm d}$, are given by \citet{2007Icar..192..588Y}:
\begin{eqnarray}
h_{\rm d} & = & h_{\rm g} {\left( 1 + \frac{{\rm St}}{\alpha_{\rm turb}} \frac{1 + 2 {\rm St}}{1 + {\rm St}} \right)}^{- 1/2}, \label{eq:hd} \\
\rho_{\rm d} & = & \frac{\Sigma_{\rm dust}}{\sqrt{2 \pi} h_{\rm d}}.
\end{eqnarray}

\subsubsection{Conversion of dust to planetesimals}
\label{sec:SI}

When the dust-to-gas mass ratio at the midplane is sufficiently high, hydrodynamic simulations revealed that part of pebbles would be converted into planetesimals via the streaming instability \citep[e.g.,][]{2015A&A...579A..43C,2017A&A...606A..80Y,2018ApJ...860..140S}\footnote{
We acknowledge that hydrodynamic simulations of planetesimal formation via the streaming instability usually assumed protoplanetary disks with negative pressure gradient \citep[e.g.,][]{2010ApJ...722L.220B,2010ApJ...722.1437B,2015A&A...579A..43C,2017A&A...606A..80Y}, and whether planetesimals can also be formed via the streaming instability in disks with positive pressure gradient is unclear.
Although a high value of $\rho_{\rm d} / \rho_{\rm g} \gtrsim 1$ should be beneficial to make planetesimals via instabilities, further studies on the condition for planetesimal formation are needed.
}.
Following the approach of \citet{2016A&A...594A.105D}, we take into account the effect of planetesimal formation.
When the midplane dust density is higher than the gas density, $\rho_{\rm d} > \rho_{\rm g}$, we convert part of dust into planetesimals as follows \citep{2016A&A...594A.105D,2019ApJ...871...10U}:
\begin{equation}
\dot{\Sigma}_{\rm plts} = 
\begin{cases}
\displaystyle - \frac{\zeta}{2 \pi} \Sigma_{\rm dust} \Omega_{\rm K} & (\rho_{\rm d} > \rho_{\rm g}), \\
0 & (\rho_{\rm d} \le \rho_{\rm g}),
\end{cases}
\label{eq:conversion}
\end{equation}
where $\zeta = 10^{-4}$ is the planetesimal formation efficiency.

In this study, we consider two types of dust particles.
As the mass loss rate of crystalline dust particles should be proportional to the crystallinity, the mass loss rate of dust particles via conversion of dust to planetesimal is given by
\begin{equation}
\dot{\Sigma}_{{\rm plts}, i} = - \frac{\zeta}{2 \pi} \Sigma_{i} \Omega_{\rm K}.
\end{equation}

\section{Results}
\label{sec:result}

In Section \ref{sec:result}, we show the results of the disk evolution and radial distribution of the crystallinity of silicate dust particles.
The results for disks without disk winds are shown in Section \ref{sec:no-wind}, and the results for evolving disks with disk winds are shown in Section \ref{sec:wind}.

\subsection{Accretion disks without disk winds}
\label{sec:no-wind}

\begin{figure}
\centering
\includegraphics[width = \columnwidth]{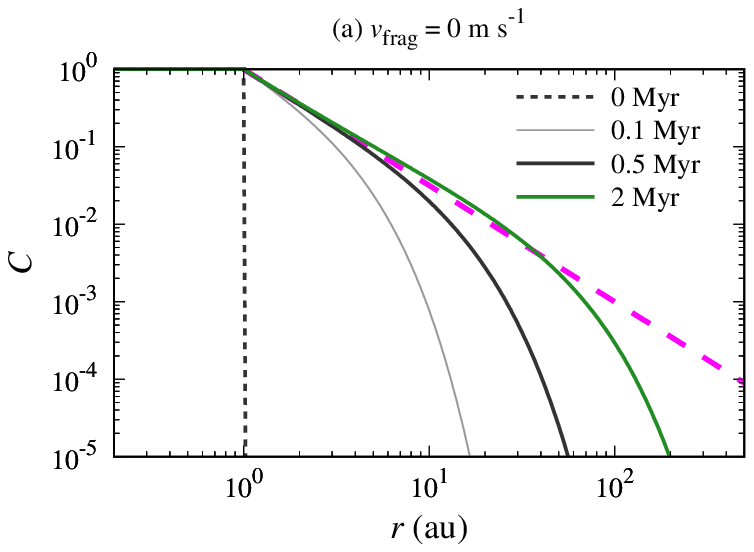}
\includegraphics[width = \columnwidth]{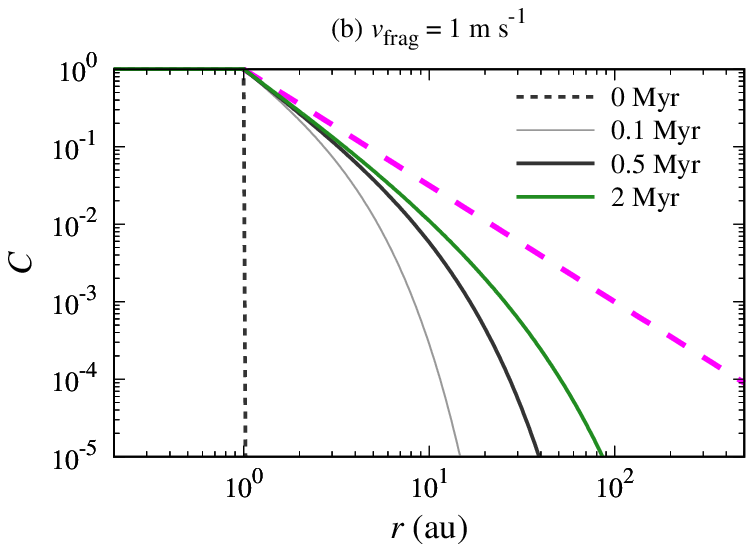}
\includegraphics[width = \columnwidth]{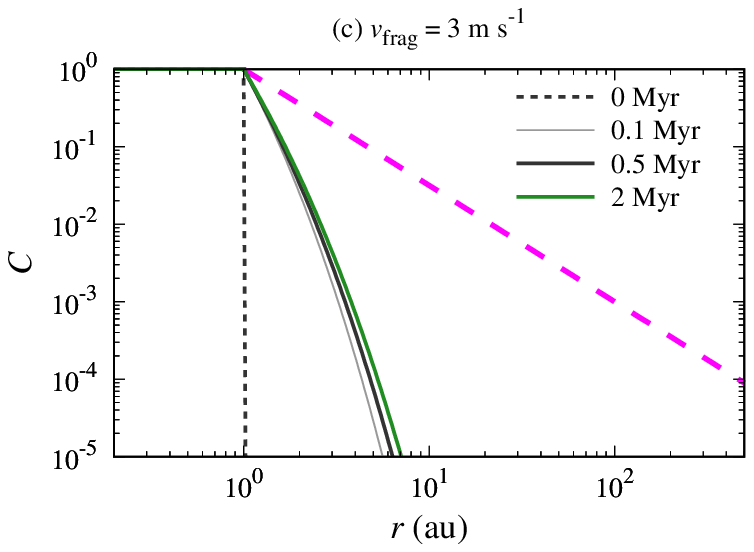}
\caption{
Radial distribution of the crystallinity.
(a) For the case of $v_{\rm frag} = 0\ {\rm m}\ {\rm s}^{-1}$ (i.e., ${\rm St} = 0$).
(b) For the case of $v_{\rm frag} = 1\ {\rm m}\ {\rm s}^{-1}$.
(c) For the case of $v_{\rm frag} = 3\ {\rm m}\ {\rm s}^{-1}$.
We set $\alpha_{\rm acc} = 10^{-3}$, $C_{\rm w} = 0$, and ${\rm Sc} = 1$.
The magenta dashed line shows the analytical solution of the radial distribution of the crystallinity in steady state disks (Equation \ref{eq:PD2007}).
}
\label{fig:C-no-wind}
\end{figure}

We performed the evolution of protoplanetary disks that evolve without disk winds.
Figure \ref{fig:C-no-wind} shows the time evolution of the radial distribution of the crystallinity of silicate dust particles.
We set $\alpha_{\rm acc} = 10^{-3}$, $C_{\rm w} = 0$, and ${\rm Sc} = 1$ in Section \ref{sec:no-wind}, and we changed the value of $v_{\rm frag}$ as a parameter.

Figure \ref{fig:C-no-wind}(a) shows the radial distribution of the crystallinity for the case of $v_{\rm frag} = 0\ {\rm m}\ {\rm s}^{-1}$ (i.e., ${\rm St} = 0$).
In this case, we found that the radial distribution at $t = 2\ {\rm Myr}$ is approximately identical to that obtained from the analytic solution for steady-state accretion disks in $r \ll r_{0}$.
Assuming that dust particles are dynamically coupled with gas (i.e., ${\rm St} = 0$), \citet{2007A&A...471..833P} derived that the radial distribution of the crystallinity in steady-state disks is given by
\begin{equation}
C = 
\begin{cases}
1 & (r \leq r_{\rm c}), \\
\displaystyle {\left( \frac{r}{r_{\rm c}} \right)}^{(3 / 2) {\rm Sc}} & (r > r_{\rm c}),
\end{cases}
\label{eq:PD2007}
\end{equation}
where $r_{\rm c} = 1\ {\rm au}$ is the location of the crystallization front.
Here we assume that the background value of the crystallinity at $r = \infty$ is zero.

\begin{figure}
\centering
\includegraphics[width = \columnwidth]{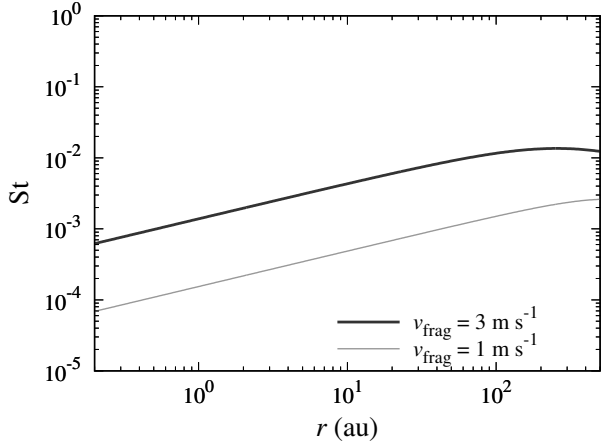}
\caption{
Radial distribution of the Stokes number at $t = 2\ {\rm Myr}$. 
We set $\alpha_{\rm acc} = 10^{-3}$, $C_{\rm w} = 0$, and ${\rm Sc} = 1$.
}
\label{fig:St-no-wind}
\end{figure}

In contrast, the radial distribution of the crystallinity is different from the analytical solution when $v_{\rm frag} \neq 0\ {\rm m}\ {\rm s}^{-1}$ (i.e., ${\rm St} \neq 0$).
Figures \ref{fig:C-no-wind}(b) and \ref{fig:C-no-wind}(c) show the radial distribution of the crystallinity for the cases of $v_{\rm frag} = 1\ {\rm m}\ {\rm s}^{-1}$ and $v_{\rm frag} = 3\ {\rm m}\ {\rm s}^{-1}$, respectively.
We found that the crystallinity decreases with increasing $v_{\rm frag}$.
The radial distribution of the Stokes number at $t = 2\ {\rm Myr}$ is shown in Figure \ref{fig:St-no-wind}.
The Stokes number increases with $r$ in most parts of the disk.
This is because the mutual collision velocity is given by ${\Delta v} \simeq {\Delta v_{\rm t}}$ and therefore $\sqrt{\rm St} c_{\rm s}$ is approximately constant (see Equations (\ref{eq:vfrag})--(\ref{eq:DeltaVt})).

The radial distribution of the crystalline silicate dust particles is given by the balance among three physical processes, that is, advection, radial drift, and diffusion of crystalline dust particles.
We define the timescales of these processes as follows:
\begin{eqnarray}
t_{\rm adv}     & = & \frac{r}{v_{\rm adv}}, \\
t_{\rm drift}   & = & \frac{r}{v_{\rm drift}}, \\
t_{\rm c, diff} & = & \frac{r}{v_{\rm c, diff}}.
\end{eqnarray}

\begin{figure}
\centering
\includegraphics[width = \columnwidth]{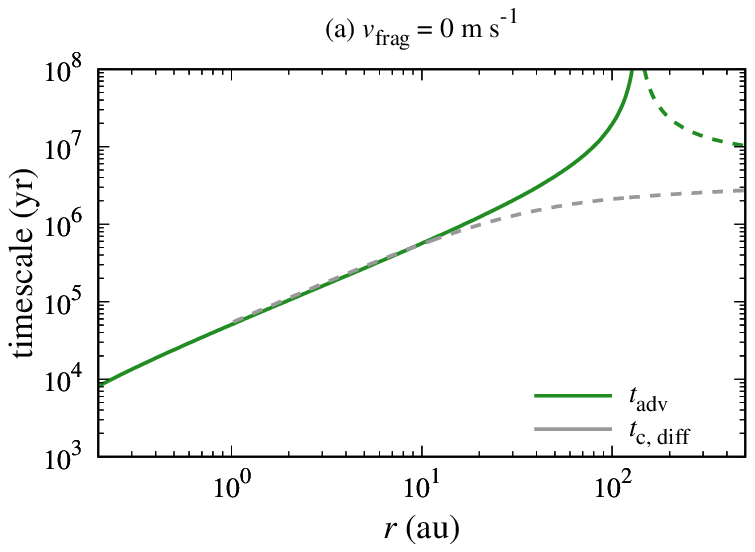}
\includegraphics[width = \columnwidth]{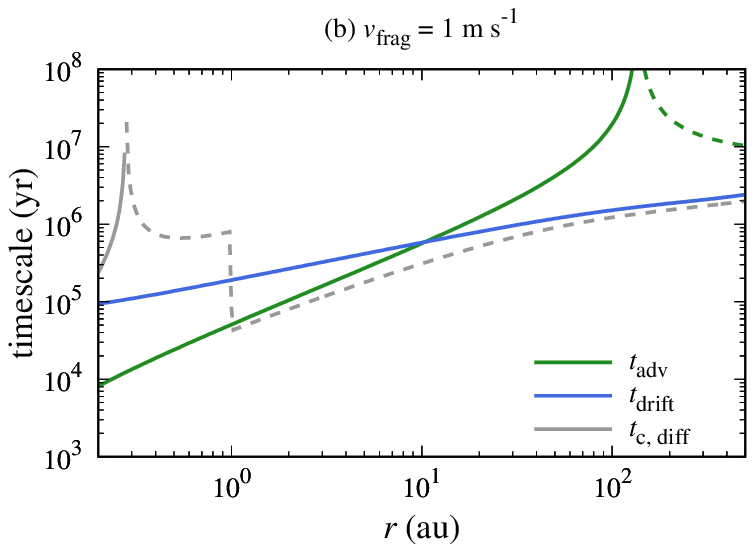}
\includegraphics[width = \columnwidth]{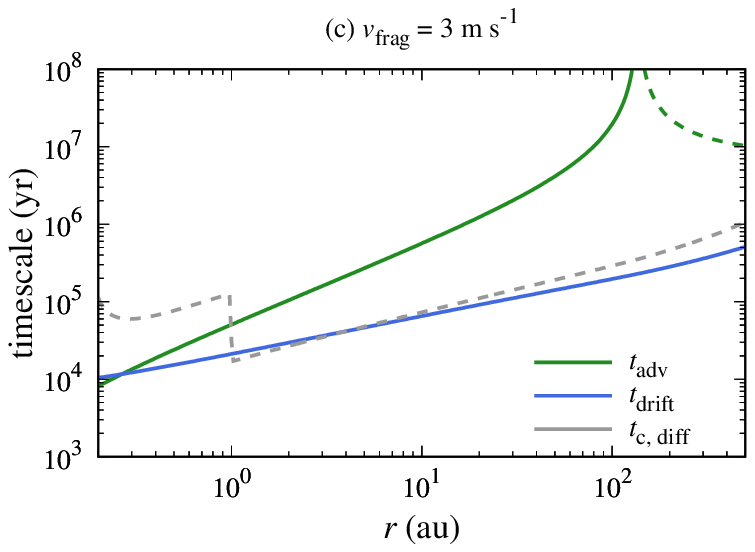}
\caption{
Timescales of advection, radial drift, and diffusion at $t = 2\ {\rm Myr}$. 
(a) For the case of $v_{\rm frag} = 0\ {\rm m}\ {\rm s}^{-1}$.
(b) For the case of $v_{\rm frag} = 1\ {\rm m}\ {\rm s}^{-1}$.
(c) For the case of $v_{\rm frag} = 3\ {\rm m}\ {\rm s}^{-1}$.
We set $\alpha_{\rm acc} = 10^{-3}$, $C_{\rm w} = 0$, and ${\rm Sc} = 1$.
Solid lines indicate that the timescales are positive, i.e., the direction of the flows is inward, and dashed lines indicate that the timescales are negative.
}
\label{fig:t-no-wind}
\end{figure}

Figure \ref{fig:t-no-wind} shows the timescales of advection, radial drift, and diffusion at $t = 2\ {\rm Myr}$.
We found that the diffusion timescale is negative in the region beyond the crystallization front ($r > r_{\rm c}$).
Therefore, the direction of the diffusion is outward.
The diffusion timescale is balanced with the advection or radial drift timescales.
In particular, the radial drift and diffusion timescales are balanced when the radial drift timescale is shorter than the advection timescale (see Figure \ref{fig:t-no-wind}(c)).

For the case of $v_{\rm frag} = 0\ {\rm m}\ {\rm s}^{-1}$, crystalline dust particles do not drift relative to the gas but diffuse due to the gradient of $\Sigma_{\rm c} / \Sigma_{\rm gas}$.
The equilibrium of the radial distribution of the crystallinity is given by the balance between diffusion and advection.
The radial distribution of the crystallinity at $t = 2\ {\rm Myr}$ is approximately consistent with the steady-state solution (Equation \ref{eq:PD2007}) at $r \ll r_{0}$.

In contrast, for the case of $v_{\rm frag} = 3\ {\rm m}\ {\rm s}^{-1}$, crystalline dust particles drift inwardly, and the effect of the inward advection is negligibly smaller than that of radial drift (Figure \ref{fig:t-no-wind}(c)).
The equilibrium of the radial distribution of the crystallinity is given by the balance between diffusion and strong radial drift.
Thus, outward transport of crystalline dust particles are suppressed compared to the case of $v_{\rm frag} = 0\ {\rm m}\ {\rm s}^{-1}$.
Figure \ref{fig:C-no-wind}(c) shows that the radial distribution of the crystallinity already reaches the steady state at $t = 2\ {\rm Myr}$.
As the radial drift timescale is inversely proportional to the Stokes number, calculations with a large value of $v_{\rm frag}$ lead to the depletion of the crystallinity beyond the crystallization front for the case of disks that evolve without disk winds.
This result is qualitatively inconsistent with the observed findings that crystalline dust particles are found farther away from the crystallization front.

\subsection{Evolving disks due to disk winds}
\label{sec:wind}

\begin{figure}
\centering
\includegraphics[width = \columnwidth]{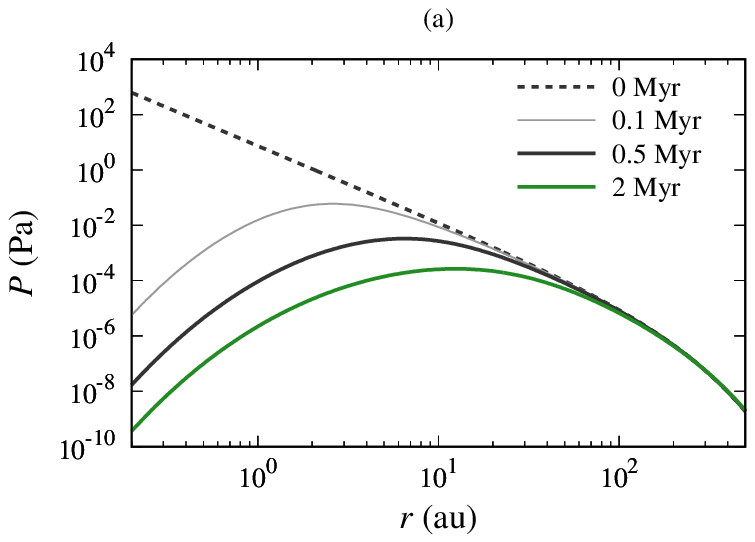}
\includegraphics[width = \columnwidth]{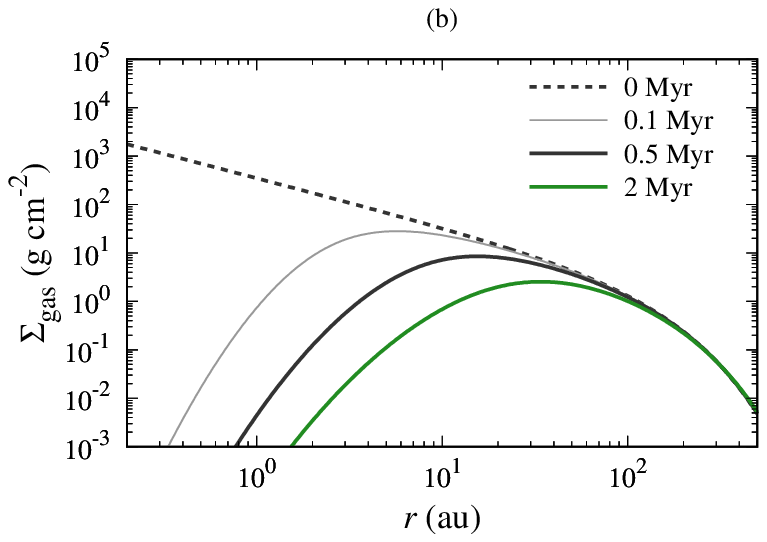}
\caption{
Radial distributions of (a) the gas pressure at the midplane and (b) the gas surface density.
We set $\alpha_{\rm acc} = 10^{-4}$, $C_{\rm w} = 10^{-5}$, and ${\rm Sc} = 1$.
}
\label{fig:gas}
\end{figure}

We performed the evolution of protoplanetary disks that evolve due to disk winds.
Figure \ref{fig:gas} shows the radial distributions of the gas pressure at the midplane and the gas surface density.
We set $\alpha_{\rm acc} = 10^{-4}$, $C_{\rm w} = 10^{-5}$, and ${\rm Sc} = 1$ in Section \ref{sec:wind}.

Figure \ref{fig:gas}(a) shows the evolution of the radial distributions of the gas pressure at the midplane.
As shown in previous studies \citep[e.g.,][]{2016A&A...596A..74S,2018ApJ...865..102T}, magnetically driven disk winds creates a maximum of gas pressure.
\citet{2018ApJ...865..102T} revealed that the location of the pressure maximum moves outward with time because the timescale of wind mass loss is longer for a larger orbital radius.
Our result is consistent with that of \citet{2018ApJ...865..102T}.
The location of the pressure maximum at $t = 2\ {\rm Myr}$ is $r = 12\ {\rm au}$, and this is approximately consistent with that obtained from the analytical solution for steady-state disks with viscous accretion and magnetically driven disk winds (see Appendix \ref{app:ss}).

Figure \ref{fig:gas}(b) also shows the evolution of the gas surface density.
It is clear that the locations of the maxima of the gas pressure and gas surface density are different: the location of the pressure maximum is inner than that of the gas surface density. 
This relation is also explained by the analytical solution for steady-state disks (see Appendix \ref{app:ss}).

\begin{figure}
\centering
\includegraphics[width = \columnwidth]{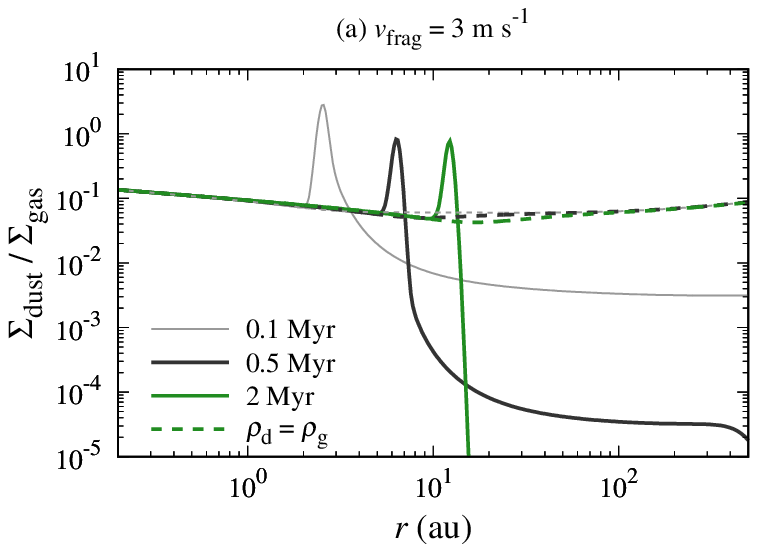}
\includegraphics[width = \columnwidth]{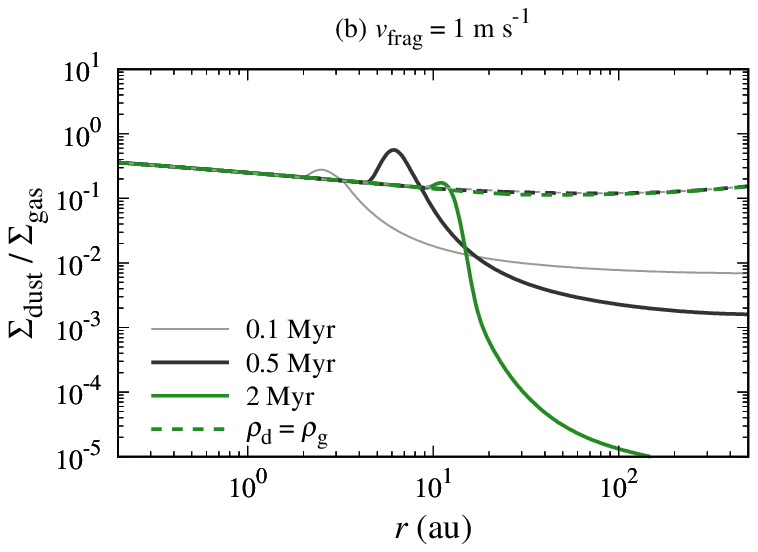}
\includegraphics[width = \columnwidth]{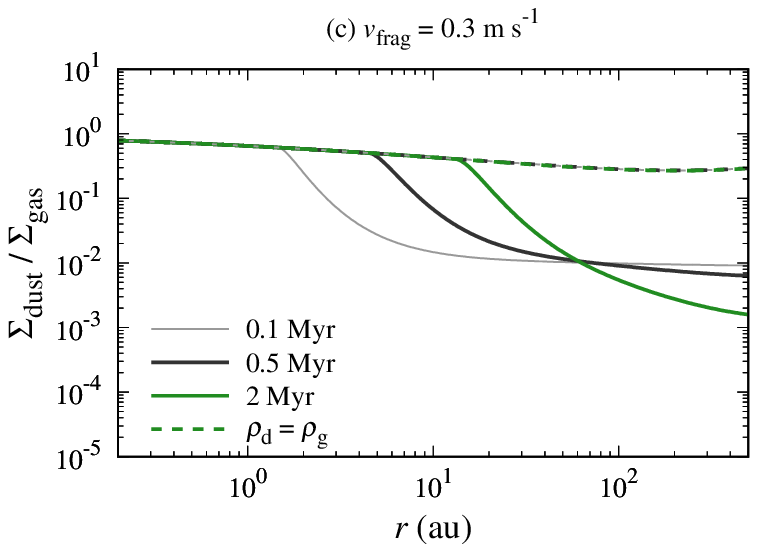}
\caption{
Radial distribution of the dust-to-gas mass ratio, $\Sigma_{\rm dust} / \Sigma_{\rm gas}$.
(a) For the case of $v_{\rm frag} = 3\ {\rm m}\ {\rm s}^{-1}$.
(b) For the case of $v_{\rm frag} = 1\ {\rm m}\ {\rm s}^{-1}$.
(c) For the case of $v_{\rm frag} = 0.3\ {\rm m}\ {\rm s}^{-1}$.
We set $\alpha_{\rm acc} = 10^{-4}$, $C_{\rm w} = 10^{-5}$, and ${\rm Sc} = 1$.
Dashed lines show the analytical estimates from Equation (\ref{eq:d-g}).
}
\label{fig:d-g}
\end{figure}

Figure \ref{fig:dust3e2}(a) shows the radial distribution of the dust surface density for the case of $v_{\rm frag} = 3\ {\rm m}\ {\rm s}^{-1}$.
We found that a narrow dust ring is formed in the disk, and the location is approximately identical to that of the pressure maximum.
This is because large pebbles are accumulated around the pressure maximum \citep[e.g.,][]{2003ApJ...583..996H,2018ApJ...865..102T}.
The dust surface density is significantly depleted beyond the dust ring due to the inward radial drift of large pebbles.
Inside the dust ring, the dust surface density is controlled by the conversion of dust to planetesimals (see Section \ref{sec:SI}).
Figure \ref{fig:d-g} shows the radial distribution of the dust-to-gas mass ratio.
We found that the dust surface density inside the dust ring is approximately given by the following equation: $\rho_{\rm d} = \rho_{\rm g}$.
This is because the conversion of dust particles to planetesimals occurs immediately when $\rho_{\rm d} > \rho_{\rm g}$ (see Equation \ref{eq:conversion}).

\begin{figure}
\centering
\includegraphics[width = \columnwidth]{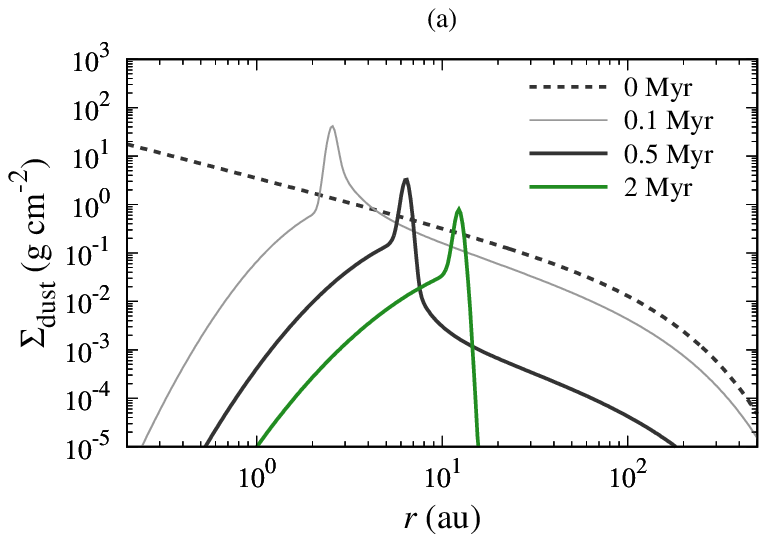}
\includegraphics[width = \columnwidth]{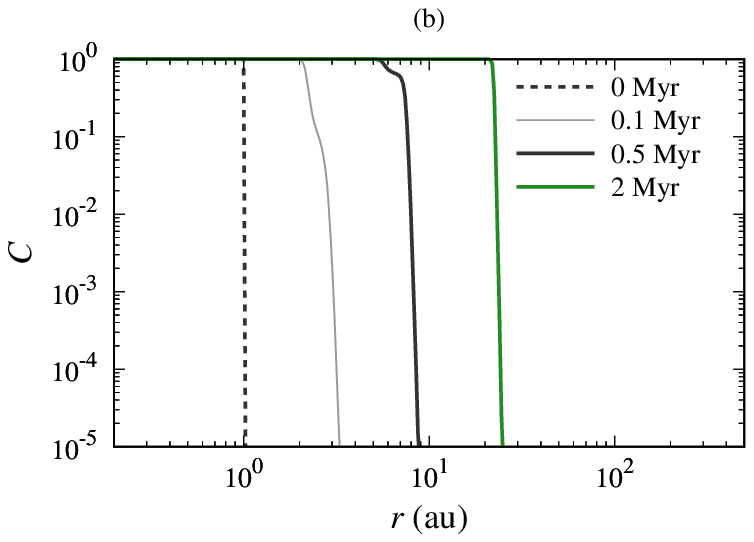}
\caption{
Radial distribution of (a) the dust surface density and (b) the crystallinity of silicate dust particles with $v_{\rm frag} = 3\ {\rm m}\ {\rm s}^{-1}$.
We set $\alpha_{\rm acc} = 10^{-4}$, $C_{\rm w} = 10^{-5}$, and ${\rm Sc} = 1$.
}
\label{fig:dust3e2}
\end{figure}

\begin{figure}
\centering
\includegraphics[width = \columnwidth]{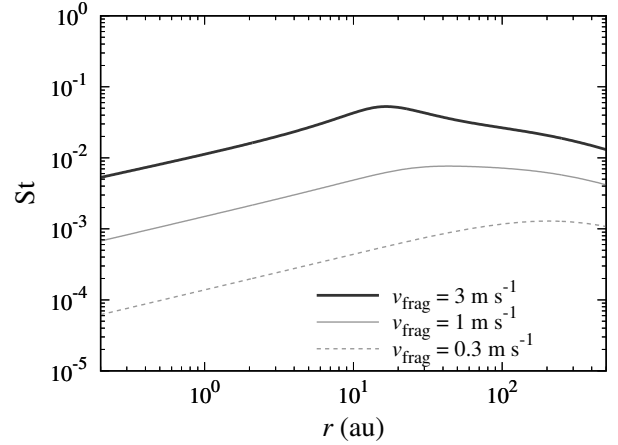}
\caption{
Radial distribution of the Stokes number at $t = 2\ {\rm Myr}$. 
}
\label{fig:St-wind}
\end{figure}

The radial distribution of the Stokes number at $t = 2\ {\rm Myr}$ is shown in Figure \ref{fig:St-wind}.
We set $\alpha_{\rm acc} = 10^{-4}$ instead of $\alpha_{\rm acc} = 10^{-3}$ in Section \ref{sec:wind}, and the Stokes number shown in Figure \ref{fig:St-wind} is larger than that shown in Figure \ref{fig:St-no-wind}.
For the case of $v_{\rm frag} = 3\ {\rm m}\ {\rm s}^{-1}$, the Stokes number is in the range of $10^{-2} \lesssim {\rm St} \lesssim 10^{-1}$ throughout the disk.
This Stokes number provides the values of the dust-to-gas mass ratio under $\rho_{\rm d} = \rho_{\rm g}$.
The dust scale height is given by Equation (\ref{eq:hd}), and the midplane dust density is inversely proportional to the dust scale height.
Therefore, $\Sigma_{\rm dust} / \Sigma_{\rm gas}$ is given by
\begin{eqnarray}
\frac{\Sigma_{\rm dust}}{\Sigma_{\rm gas}} & = & {\left( 1 + \frac{{\rm St}}{\alpha_{\rm turb}} \frac{1 + 2 {\rm St}}{1 + {\rm St}} \right)}^{- 1/2}, \label{eq:d-g} \\
& \simeq & \sqrt{\frac{\alpha_{\rm turb}}{{\rm St}}},
\end{eqnarray}
when $\rho_{\rm d} = \rho_{\rm g}$ is achieved.
This estimation explains the radial distribution of $\Sigma_{\rm dust} / \Sigma_{\rm gas}$ shown in Figure \ref{fig:d-g}.

Figure \ref{fig:dust3e2}(b) shows the radial distribution of the crystallinity of silicate dust particles for the case of $v_{\rm frag} = 3\ {\rm m}\ {\rm s}^{-1}$.
We found that the crystallinity is almost 100\% around and inside the location of the dust ring.
The radial distribution of the crystallinity shown in Figure \ref{fig:dust3e2}(b) is completely different from that for disks without disk winds.
In Section \ref{sec:new-mechanism}, we unveil the mechanism for the radial transport of crystalline dust particles in evolving disk due to disk wind.
The key physics of the efficient radial transport is the {\it outward radial drift} of pebbles.

\begin{figure}
\centering
\includegraphics[width = \columnwidth]{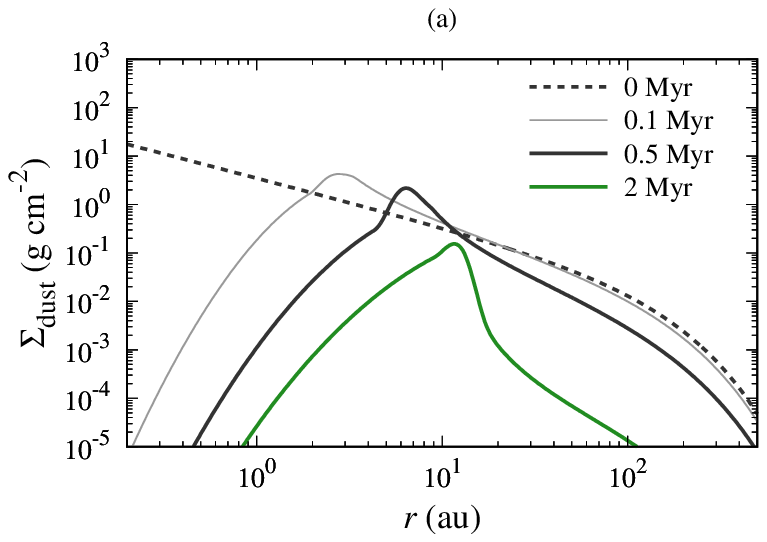}
\includegraphics[width = \columnwidth]{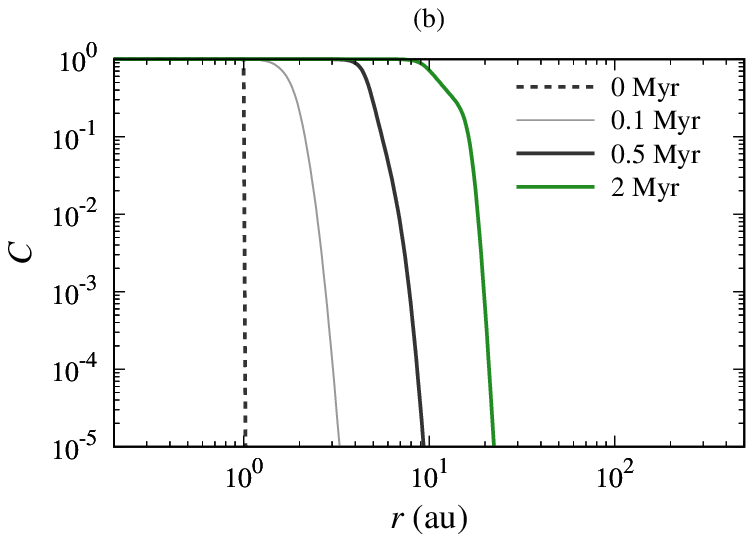}
\caption{
Radial distribution of (a) the dust surface density and (b) the crystallinity of silicate dust particles with $v_{\rm frag} = 1\ {\rm m}\ {\rm s}^{-1}$.
We set $\alpha_{\rm acc} = 10^{-4}$, $C_{\rm w} = 10^{-5}$, and ${\rm Sc} = 1$.
}
\label{fig:dust1e2}
\end{figure}

\begin{figure}
\centering
\includegraphics[width = \columnwidth]{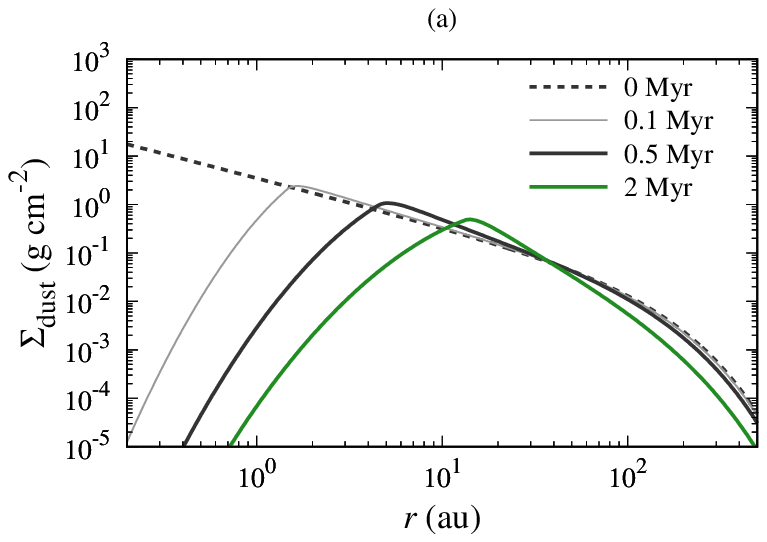}
\includegraphics[width = \columnwidth]{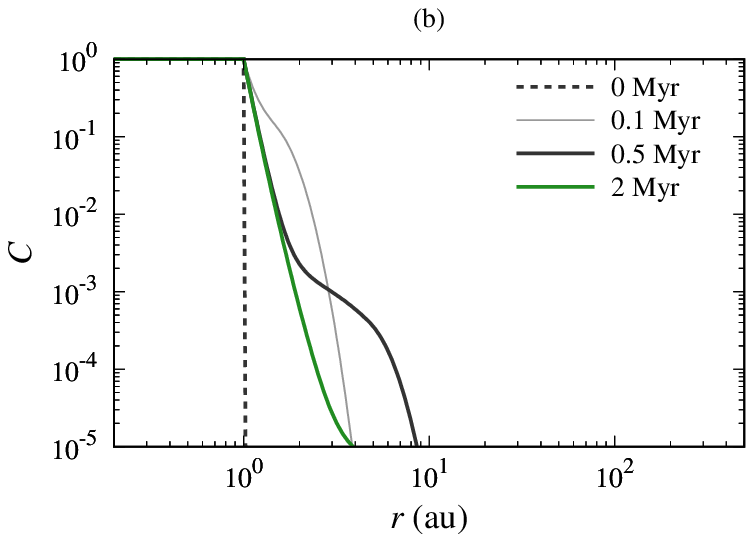}
\caption{
Radial distribution of (a) the dust surface density and (b) the crystallinity of silicate dust particles with $v_{\rm frag} = 0.3\ {\rm m}\ {\rm s}^{-1}$.
We set $\alpha_{\rm acc} = 10^{-4}$, $C_{\rm w} = 10^{-5}$, and ${\rm Sc} = 1$.
}
\label{fig:dust3e1}
\end{figure}

Figures \ref{fig:dust1e2} and \ref{fig:dust3e1} show the radial distributions of $\Sigma_{\rm dust}$ and $C$ for the cases of $v_{\rm frag} = 1\ {\rm m}\ {\rm s}^{-1}$ and $v_{\rm frag} = 0.3\ {\rm m}\ {\rm s}^{-1}$, respectively.
As shown in Figure \ref{fig:dust1e2}(a), the radial distribution of the dust surface density for the case of $v_{\rm frag} = 1\ {\rm m}\ {\rm s}^{-1}$ is similar to that for the case of $v_{\rm frag} = 3\ {\rm m}\ {\rm s}^{-1}$, although the width and the maximum value of $\Sigma_{\rm dust}$ of the dust ring are different.
The radial distribution of the crystallinity shown in Figure \ref{fig:dust1e2}(b) is also similar to that shown in Figure \ref{fig:dust3e2}(b).

In contrast, the radial distributions of $\Sigma_{\rm dust}$ and $C$ for the cases of $v_{\rm frag} = 0.3\ {\rm m}\ {\rm s}^{-1}$ are completely different from those for $v_{\rm frag} = 3\ {\rm m}\ {\rm s}^{-1}$.
As shown in Figure \ref{fig:dust3e1}(a), the dust surface density hardly changes with time beyond the maximum of the dust surface density.
On the other hand, inside the maximum of the dust surface density, the dust-to-gas mass ratio is controlled by Equation (\ref{eq:d-g}).
Then, the dust surface density is approximately given by the following equation for the case of small pebbles: $\Sigma_{\rm dust} \simeq \min{\left( \sqrt{\alpha_{\rm turb}/{\rm St}} \Sigma_{\rm gas}, \Sigma_{{\rm dust}, 0} \right)}$.
The location of the dust ring is therefore not necessarily identical to that of the pressure maximum.

The radial distribution of the crystallinity of silicate dust particles with $v_{\rm frag} = 0.3\ {\rm m}\ {\rm s}^{-1}$ is shown in Figure \ref{fig:dust3e1}(b).
In contrast to the radial distribution shown in Figures \ref{fig:dust3e2}(b) and \ref{fig:dust1e2}(b), the crystallinity is $C < 1$ outside the crystallization front, and the crystallinity decreases with increasing $r$.  
The crystallinity around the dust ring is $C \ll 1$ for the case of $v_{\rm frag} = 0.3\ {\rm m}\ {\rm s}^{-1}$.

\subsection{Outward radial drift as a new mechanism for radial transport of crystalline dust particles}
\label{sec:new-mechanism}

As shown in Figures \ref{fig:dust3e2}(b) and \ref{fig:dust1e2}(b), the crystallinity is almost 100\% around and inside the location of the dust ring when the threshold velocity for collisional fragmentation/growth is $v_{\rm frag} \ge 1\ {\rm m}\ {\rm s}^{-1}$.
In Section \ref{sec:new-mechanism}, we show the condition for driving efficient radial transport.

\begin{figure}
\centering
\includegraphics[width = \columnwidth]{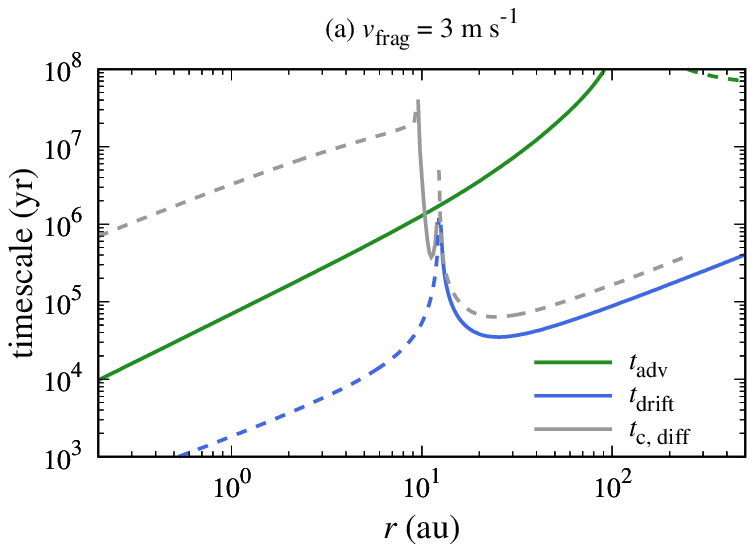}
\includegraphics[width = \columnwidth]{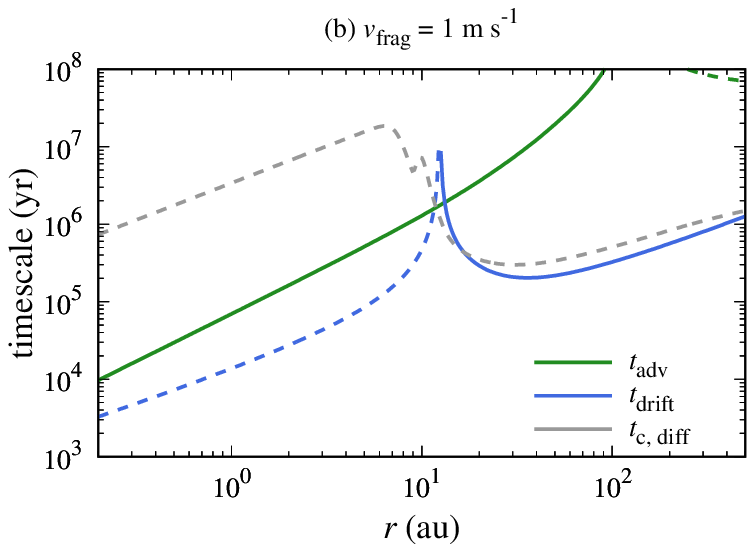}
\includegraphics[width = \columnwidth]{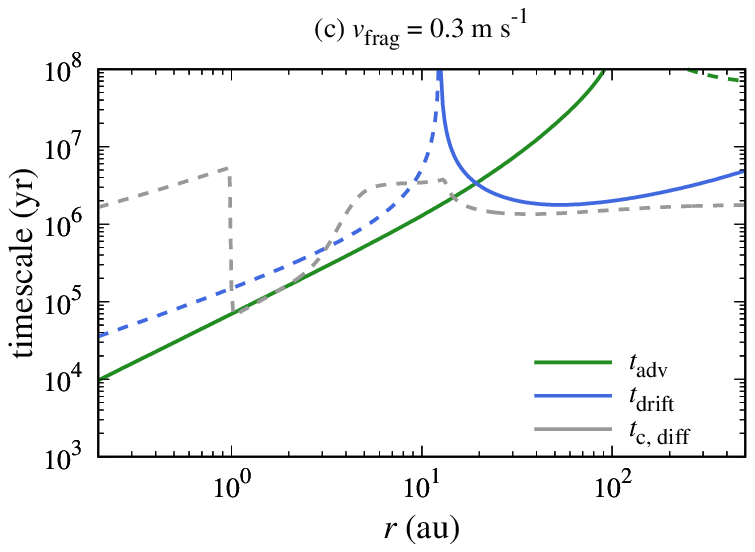}
\caption{
Timescales of advection, radial drift, and diffusion at $t = 2\ {\rm Myr}$. 
(a) For the case of $v_{\rm frag} = 3\ {\rm m}\ {\rm s}^{-1}$.
(b) For the case of $v_{\rm frag} = 1\ {\rm m}\ {\rm s}^{-1}$.
(c) For the case of $v_{\rm frag} = 0.3\ {\rm m}\ {\rm s}^{-1}$.
We set $\alpha_{\rm acc} = 10^{-4}$, $C_{\rm w} = 10^{-5}$, and ${\rm Sc} = 1$.
}
\label{fig:t-wind}
\end{figure}

Figure \ref{fig:t-wind} shows the timescales of advection, radial drift, and diffusion at $t = 2\ {\rm Myr}$.
Inside the pressure maximum, the radial drift timescale is negative while the advection timescale is positive.
For the case of $v_{\rm frag} \ge 1\ {\rm m}\ {\rm s}^{-1}$, the Stokes number of pebbles is large and the radial drift timescale is shorter than the advection timescale: ${\left| t_{\rm drift} \right|} < {\left| t_{\rm adv} \right|}$ (see Figures \ref{fig:t-wind}(a) and \ref{fig:t-wind}(b)).
In this case, the outward radial drift of pebbles can transport the crystalline dust particles from the crystallization front to the pressure maximum.
Then, the crystallinity reaches almost 100\% around and inside the location of the dust ring.

In contrast, for the case of $v_{\rm frag} = 0.3\ {\rm m}\ {\rm s}^{-1}$, the Stokes number of pebbles is small and the radial drift timescale is longer than the advection timescale: ${\left| t_{\rm drift} \right|} > {\left| t_{\rm adv} \right|}$ (see Figure \ref{fig:t-wind}(c)).
In this case, the outward radial drift of pebbles {\it cannot} transport the crystalline dust particles efficiently.
Then, the diffusion timescale is balanced with the advection timescale, and the crystallinity decreases with increasing $r$.

\begin{figure*}
\centering
\includegraphics[width = 0.8\textwidth]{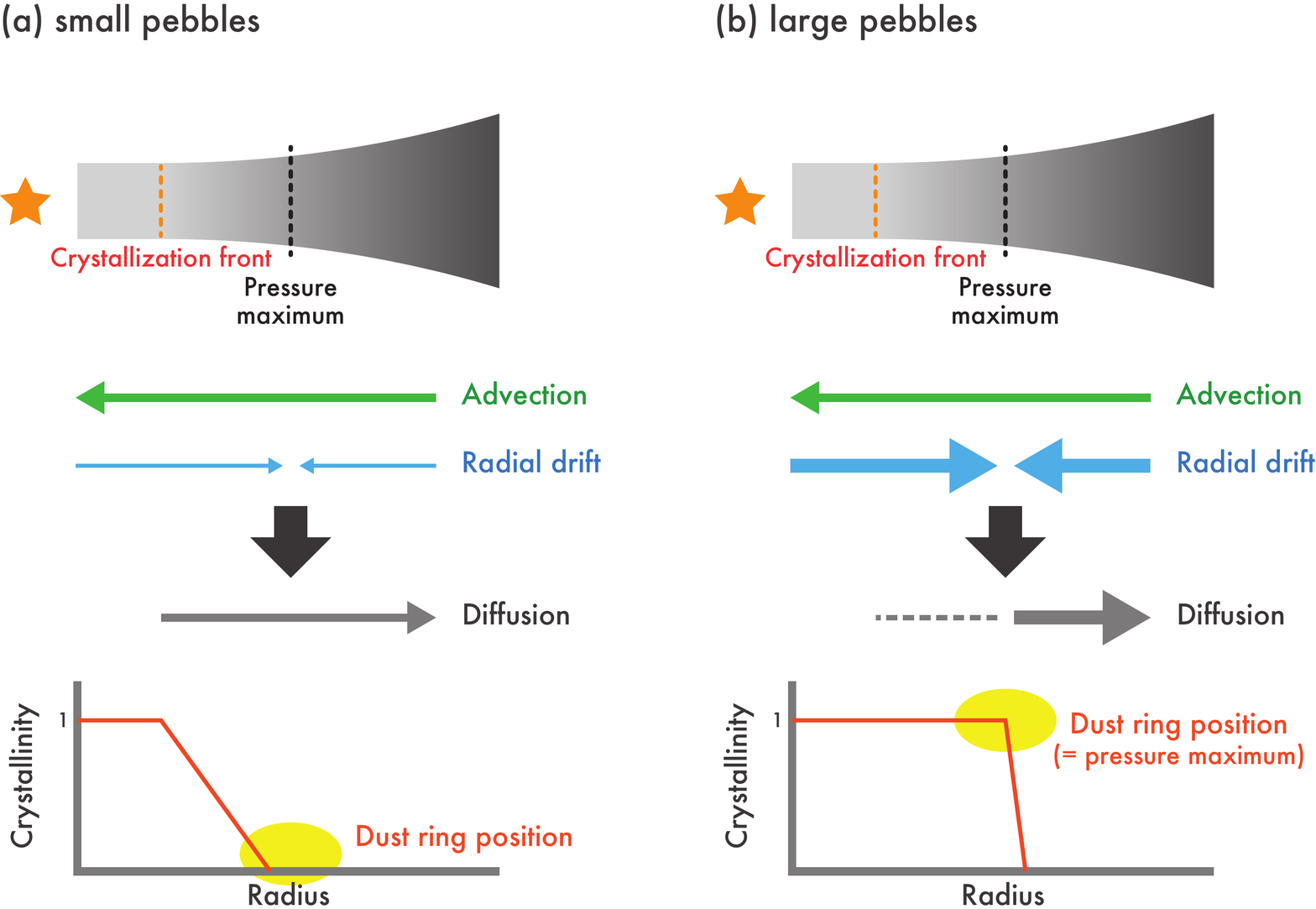}
\caption{
Schematic of the radial transport processes in evolving protoplanetary disks due to magnetically driven disk winds.
(a) For the case of small pebbles, i.e., ${\left| t_{\rm drift} \right|} > {\left| t_{\rm adv} \right|}$.
The outward radial drift of pebbles cannot transport the crystalline dust particles, and the diffusion timescale is balanced with the advection timescale.
(b) For the case of large pebbles, i.e., ${\left| t_{\rm drift} \right|} < {\left| t_{\rm adv} \right|}$.
If the radial drift timescale is shorter than the advection timescale, the {\it outward radial drift} of pebbles can transport the crystalline dust particles from the crystallization front to the pressure maximum efficiently.
}
\label{fig:sch}
\end{figure*}

Our novel mechanism for the radial transport of crystalline dust particles is illustrated in Figure \ref{fig:sch}.
The condition for driving efficient radial transport by the outward radial drift is ${\left| t_{\rm drift} \right|} < {\left| t_{\rm adv} \right|}$.
As the radial drift timescale is inversely proportional to the Stokes number, calculations with a large value of $v_{\rm frag}$ lead to the efficient radial transport.
Therefore, we expect that the crystallinity around and inside the dust ring reflects the size of pebbles and the threshold velocity for collisional fragmentation/growth.

By comparing our calculations with the observational findings that the crystalline silicate dust particles exist in the cold regions of the protoplanetary disks and the solar nebula, we suggest that this outward radial drift would be the key mechanism to transport crystalline silicate dust particles.
This idea is supported by the fact that the recent high-spatial-resolution observations revealed that the ring structures are relatively common among protoplanetary disks.
As the radial structures of gas disks strongly affect the dynamics of pebbles, further observational studies on the link between disk structure and dust composition are required.

\section{Dependence on the Schmidt number}
\label{sec:Sc}

In Section \ref{sec:result}, we set ${\rm Sc} = 1$ for simplicity.
However, the Schmidt number of protoplanetary disks does not necessarily have to be ${\rm Sc} = 1$.
Three-dimensional magneto-hydrodynamic simulations indicated that the turbulent diffusion due to magneto-rotational instability is expressed by the Schmidt number with $0.85 \lesssim {\rm Sc} \lesssim 10$ \citep{2005MNRAS.358.1055C,2005ApJ...634.1353J}.
Based on analytic arguments, \citet{2007A&A...471..833P} also derived the theoretical minimum value of ${\rm Sc} = 1 / 3$.
In Section \ref{sec:Sc}, we briefly review the dependence of the radial distribution of the crystalline dust particles on the Schmidt number.

\subsection{Accretion disks without disk winds}
\label{sec:Sc-no-wind}

\begin{figure}
\centering
\includegraphics[width = \columnwidth]{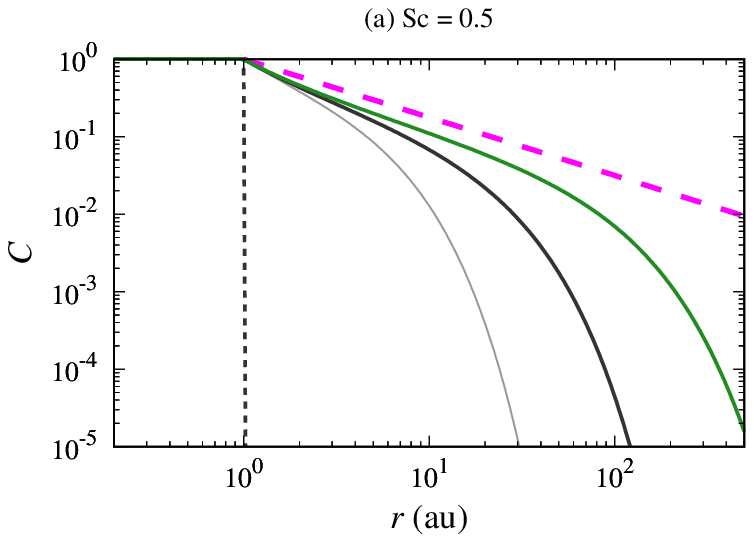}
\includegraphics[width = \columnwidth]{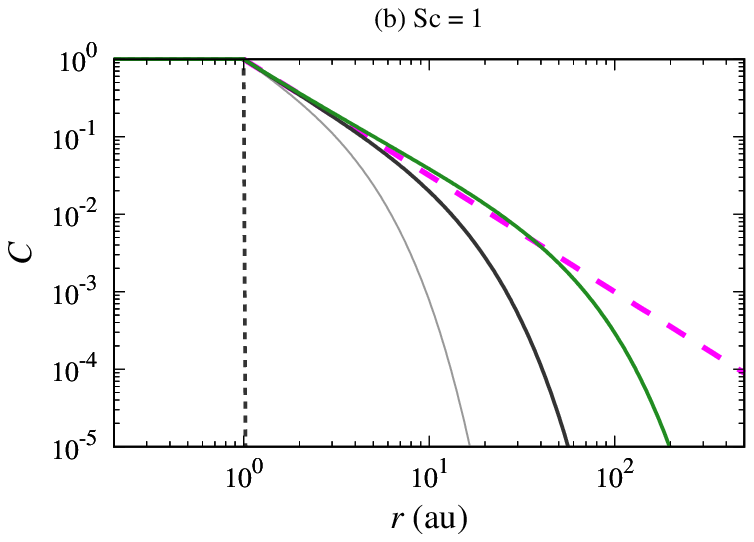}
\includegraphics[width = \columnwidth]{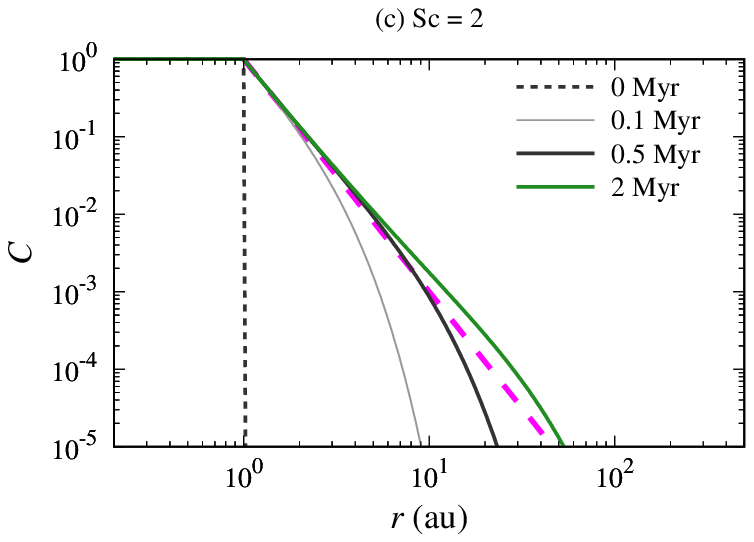}
\caption{
Radial distribution of the crystallinity.
(a) For the case of ${\rm Sc} = 0.5$.
(b) For the case of ${\rm Sc} = 1$.
(c) For the case of ${\rm Sc} = 2$.
We set $\alpha_{\rm acc} = 10^{-3}$, $C_{\rm w} = 0$, and $v_{\rm frag} = 0\ {\rm m}\ {\rm s}^{-1}$.
The magenta dashed line shows the analytical solution of the radial distribution of the crystallinity in steady state disks (Equation \ref{eq:PD2007}).
}
\label{fig:C-Sc-no-wind}
\end{figure}

Here we show the results for accretion disks without disk winds in Section \ref{sec:Sc-no-wind}.
Figure \ref{fig:C-Sc-no-wind} shows the time evolution of the radial distribution of the crystallinity of silicate dust particles.
Here we set $\alpha_{\rm acc} = 10^{-3}$, $C_{\rm w} = 0$, and $v_{\rm frag} = 0\ {\rm m}\ {\rm s}^{-1}$, and we changed the value of ${\rm Sc}$ as a parameter.

We confirmed that the radial distribution at $t = 2\ {\rm Myr}$ is approximately identical to that obtained from the analytical solution for steady-state accretion disks: $C = {( r / r_{\rm c} )}^{(3/2) {\rm Sc}}$ \citep{2007A&A...471..833P}.
Therefore, the radial distribution of the crystallinity is a sensitive function of ${\rm Sc}$ for the case of classical accretion disks without disk winds as shown in previous studies \citep[e.g.,][]{1988MNRAS.235..365C,2007A&A...471..833P}.
It should be noted that these radial distributions of the crystallinity are derived under $v_{\rm frag} = 0\ {\rm m}\ {\rm s}^{-1}$.
The radial distribution of the crystallinity is determined by the balance among the advection, diffusion, and radial drift when $v_{\rm frag} \neq 0\ {\rm m}\ {\rm s}^{-1}$ (Section \ref{sec:no-wind}).

\subsection{Evolving disks due to disk winds}
\label{sec:Sc-wind}

\begin{figure}
\centering
\includegraphics[width = \columnwidth]{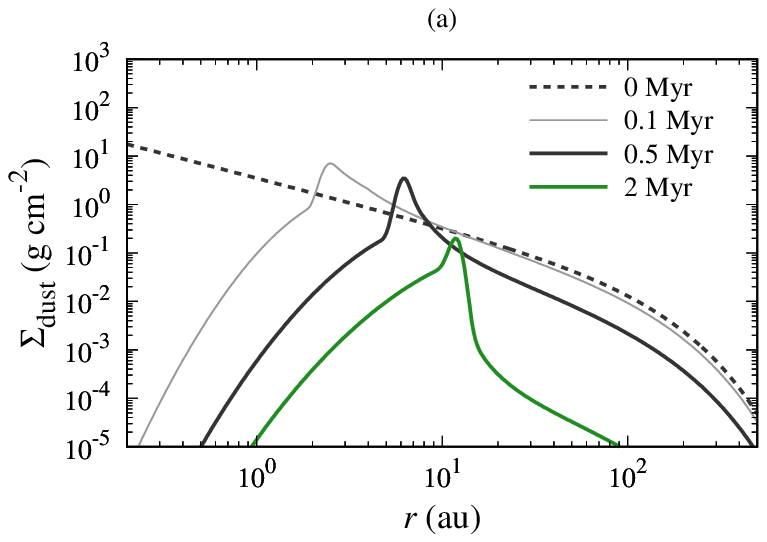}
\includegraphics[width = \columnwidth]{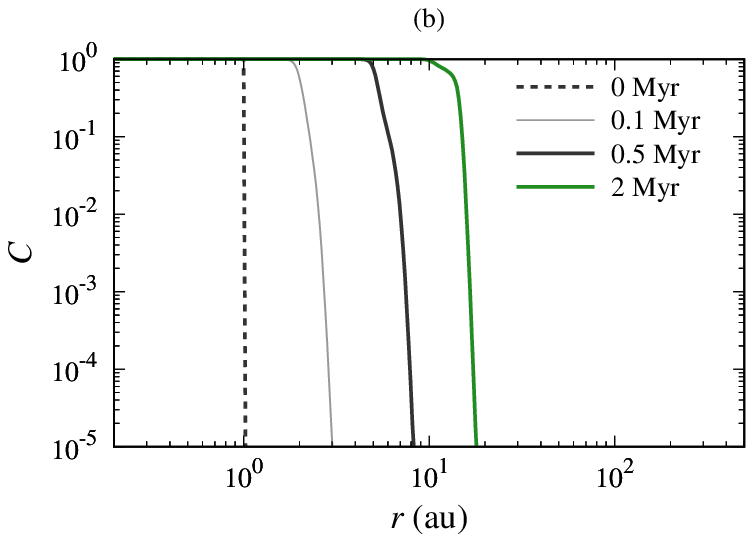}
\caption{
Radial distribution of (a) the dust surface density and (b) the crystallinity of silicate dust particles.
We set $\alpha_{\rm acc} = 10^{-4}$, $C_{\rm w} = 10^{-5}$, $v_{\rm frag} = 1\ {\rm m}\ {\rm s}^{-1}$, and ${\rm Sc} = 2$.
}
\label{fig:dust-Sc}
\end{figure}

In contrast, the radial distribution of the crystallinity is {\it not} a sensitive function of ${\rm Sc}$ for the case of evolving disks due to disk winds.
We set $\alpha_{\rm acc} = 10^{-4}$, $C_{\rm w} = 10^{-5}$, $v_{\rm frag} = 1\ {\rm m}\ {\rm s}^{-1}$, and ${\rm Sc} = 2$ in Section \ref{sec:Sc-wind}.
Figure \ref{fig:dust-Sc}(a) shows the radial distribution of the dust surface density.
A dust ring is formed around the pressure maximum as in the case of Figures \ref{fig:dust3e2}(a) and \ref{fig:dust1e2}(a).
Figure \ref{fig:dust-Sc}(b) shows the radial distribution of the crystallinity of silicate dust particles.
The crystallinity is almost 100\% around and inside the location of the dust ring as in the case of Figures \ref{fig:dust3e2}(b) and \ref{fig:dust1e2}(b).

\begin{figure}
\centering
\includegraphics[width = \columnwidth]{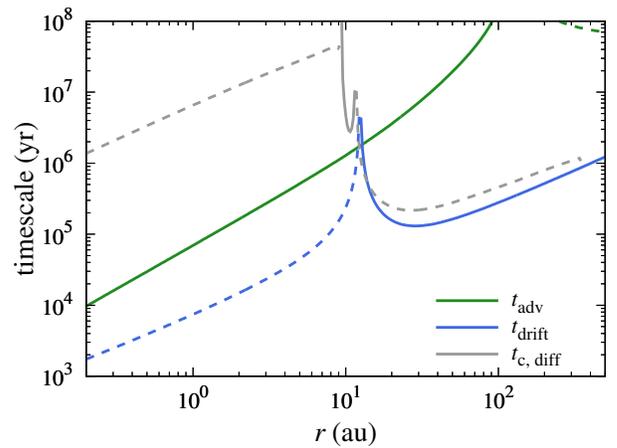}
\caption{
Timescales of advection, radial drift, and diffusion at $t = 2\ {\rm Myr}$. 
We set $\alpha_{\rm acc} = 10^{-4}$, $C_{\rm w} = 10^{-5}$, $v_{\rm frag} = 1\ {\rm m}\ {\rm s}^{-1}$, and ${\rm Sc} = 2$.
}
\label{fig:t-Sc-wind}
\end{figure}

Figure \ref{fig:t-Sc-wind} shows the timescales of advection, radial drift, and diffusion at $t = 2\ {\rm Myr}$.
In this case, the Stokes number of pebbles is large enough to satisfy the following condition: ${\left| t_{\rm drift} \right|} < {\left| t_{\rm adv} \right|}$.
Then, the outward radial drift of pebbles can transport the crystalline dust particles from the crystallization front to the pressure maximum, and the crystallinity reaches almost 100\% around and inside the location of the dust ring.

For the case of classical accretion disks without disk winds, the radial distribution of the crystallinity is given by the balance of the diffusion and advection (or radial drift) timescales.
In contrast, for the case of evolving disks due to disk winds, the radial distribution of the crystallinity is almost 100\% if the outward radial drift overcomes the inward advection.
As the diffusion is not the main mechanism, the radial distribution of the crystallinity hardly depends on ${\rm Sc}$ as long as the condition for the outward radial drift (i.e., ${\left| t_{\rm drift} \right|} < {\left| t_{\rm adv} \right|}$) is satisfied (see Figure \ref{fig:sch}).

\section{Summary}

Several pieces of evidence suggest that silicate dust particles undergo significant thermal processing in protoplanetary disks, and crystalline dust particles should transport outwardly as they are found in the outer region of protoplanetary disks and the solar system.
Several mechanisms have been proposed for the outward transport of crystalline dust particles \citep[e.g.,][]{2001A&A...378..192G,2004A&A...415.1177K,2006ApJ...640L..67D,2010Icar..208..455C,2012M&PS...47...99Y}.

Recent astronomical observations revealed varieties of structures of protoplanetary disks.
In particular, the dust ring structures are observed in a large number of disks, and the accumulation of pebbles around the pressure maximum created by mass loss due to magnetically driven disk winds is one of the possible origins of the observed dust ring structure \citep[e.g.,][]{2018ApJ...865..102T}.

In this study, we proposed a novel mechanism for the outward transport of crystalline dust particles.
In evolving protoplanetary disks due to magnetically driven disk winds, dust particles can migrate outwardly by radial drift.
We found that the outward radial drift process can transport crystalline dust particles efficiently when the radial drift overcomes the advective flow.
Our findings are summarized as follows.

\begin{enumerate}
\item{
In Section \ref{sec:no-wind}, we performed the evolution of protoplanetary disks that evolve without disk winds.
The diffusion timescale is balanced with the advection or inward radial drift timescales (see Figure \ref{fig:t-no-wind}); the crystallinity is well expressed by the analytical estimation by \cite{2007A&A...471..833P} when the advection balances with the diffusion.
It should be noted that the inward radial drift significantly suppresses the outward transport of the crystalline dust particles, which is inconsistent with the observational evidence.
}
\item{
In Section \ref{sec:wind}, we performed the evolution of protoplanetary disks that evolve due to disk winds.
As shown in previous studies \citep[e.g.,][]{2016A&A...596A..74S,2018ApJ...865..102T}, magnetically driven disk winds create a maximum of gas pressure at a certain radius (see Figure \ref{fig:gas}).
We found that the location of the pressure maximum at $t = 2\ {\rm Myr}$ is approximately consistent with that obtained from the analytical solution for steady-state disks with viscous accretion and magnetically driven disk winds (see Appendix \ref{app:ss}).
}
\item{
Figure \ref{fig:dust3e2}(a) shows the radial distribution of the dust surface density.
For the case of $v_{\rm frag} = 3\ {\rm m}\ {\rm s}^{-1}$, a narrow dust ring is formed in the disk, and the location is approximately identical to that of the pressure maximum.
This is because large pebbles are accumulated around the pressure maximum \citep[e.g.,][]{2003ApJ...583..996H,2018ApJ...865..102T}.
}
\item{
Figure \ref{fig:dust3e2}(b) shows the radial distribution of the crystallinity of silicate dust particles.
For the case of $v_{\rm frag} = 3\ {\rm m}\ {\rm s}^{-1}$, the crystallinity is almost 100\% around and inside the location of the dust ring.
We proposed that the key physics of the efficient outward radial transport is the outward radial drift of pebbles.
}
\item{
The mechanism for the radial transport of crystalline dust particles proposed in this study is illustrated in Figure \ref{fig:sch}.
The condition for driving efficient radial transport by the outward radial drift is ${\left| t_{\rm drift} \right|} < {\left| t_{\rm adv} \right|}$.
As the radial drift timescale is inversely proportional to the Stokes number, calculations with a large value of $v_{\rm frag}$ lead to the efficient radial transport.
Therefore, we expect that the crystallinity around and inside the dust ring reflects the size of pebbles and the threshold velocity for collisional fragmentation/growth.
}
\end{enumerate}

\acknowledgments

We thank Shoji Mori, Taishi Nakamoto, and Tetsuo Taki for useful discussions.
S.A.\ was supported by JSPS KAKENHI Grant No.\ JP20J00598.
This work was supported by the Publications Committee of NAOJ.

%\software{}

%\clearpage
\appendix

\section{Stationary solution for gas disk}
\label{app:ss}

We found a stationary solution for the gas surface density of disk with viscous accretion and magnetically driven disk winds.
We set the radial distribution of the midplane temperature and the sound velocity as follows:
\begin{eqnarray}
T & = & T_{1} {\tilde{r}}^{-q}, \\
c_{\rm s} & = & c_{{\rm s}, 1} {\tilde{r}}^{- q/2},
\end{eqnarray}
where ${\tilde{r}} \equiv r / {( 1\ {\rm au} )}$ is the normalized distance from the central star, and $c_{{\rm s}, 1} = {\left( k_{\rm B} T_{1} / m_{\rm g} \right)}^{1/2}$ is the sound speed at $r = 1\ {\rm au}$.
Here $k_{\rm B}$ is the Boltzmann constant, and $m_{\rm g} = 2.34 m_{\rm H}$ is the mean molecular mass, where $m_{\rm H}$ is the mass of a hydrogen atom.

Assuming that a gas disk is in steady state, the left-hand side of Equation (\ref{eq:Sigmag}) is zero:
\begin{equation}
0 = \frac{1}{2 \pi r} \frac{\partial \dot{M}_{\rm gas}}{\partial r} - C_{\rm w} \Sigma_{\rm gas} \Omega_{\rm K}.
\end{equation}
We can rewrite the above equation as follows:
\begin{eqnarray}
{\tilde{r}}^{3 - q} \frac{\partial^{2} \Sigma_{\rm gas}}{\partial {\tilde{r}}^{2}} + {\left( \frac{9}{2} - 2 q \right)} {\tilde{r}}^{2 - q} \frac{\partial \Sigma_{\rm gas}}{\partial \tilde{r}} && \nonumber \\
+ {\left[ {\left( 2 - q \right)} {\left( \frac{3}{2} - q \right)} {\tilde{r}}^{1 - q} - \mathcal{A} \right]} \Sigma_{\rm gas} & = & 0,
\label{eq:App1}
\end{eqnarray}
where the dimensionless parameter $\mathcal{A}$ is
\begin{eqnarray}
\mathcal{A} & = & \frac{C_{\rm w} {v_{{\rm K}, 1}}^{2}}{3 \alpha_{\rm acc} {c_{{\rm s}, 1}}^{2}} \nonumber \\
& = & 26.1 {\left( \frac{C_{\rm w} / \alpha_{\rm acc}}{10^{-1}} \right)} {\left( \frac{T_{1}}{800\ {\rm K}} \right)}^{-1} {\left( \frac{M_{\star}}{2.5 M_{\odot}} \right)},
\label{eq:App2}
%26.101752083
\end{eqnarray}
where $v_{{\rm K}, 1} = \sqrt{G M_{\star} / {\left( 1\ {\rm au} \right)}}$ is the Kepler velocity at $r = 1\ {\rm au}$.

We found that the solution of Equation (\ref{eq:App1}) is given as follows:
{\small
\begin{equation}
\Sigma_{\rm gas} = \Sigma_{0} {\tilde{r}}^{q - 7/4} \frac{2 {\left( 2 p {\mathcal{A}}^{1/2} \right)}^{p}}{\Gamma {\left( p \right)}} {K_{p} {\left( 4 p {\mathcal{A}}^{1/2} {\tilde{r}}^{- 1 / {\left( 4 p \right)}} \right)}},
\end{equation}
}where $K_{p} {( x )}$ is the modified Bessel function of the second kind, and the exponent, $p$, is given by
\begin{equation}
p \equiv \frac{1}{2 {\left( 1 - q \right)}}.
\end{equation}

For the special case of $q = 1/2$ and $p = 1$, the stationary solution of $\Sigma_{\rm gas}$ for a gas disk with radial mass accretion and wind-driven mass loss is given by
\begin{equation}
\Sigma_{\rm gas} = \Sigma_{0} {\tilde{r}}^{- 5/4} \cdot 4 {\mathcal{A}}^{1/2} {K_{1} {\left( 4 {\mathcal{A}}^{1/2} {\tilde{r}}^{- 1/4} \right)}},
\label{eq:steadySigma}
\end{equation}
where $\Sigma_{0}$ is a parameter.
In this case, $\Sigma_{\rm gas}$ takes the maximum at
\begin{equation}
r = 0.66 {\mathcal{A}}^{2}\ {\rm au}.
%0.66451858117124900459
\end{equation}

We can also calculate the radial profile of the gas pressure at the midplane.
The gas pressure is given by
{\footnotesize
\begin{eqnarray}
P & = & \frac{\Sigma_{\rm gas} c_{\rm s} \Omega_{\rm K}}{\sqrt{2 \pi}} \nonumber \\
  & = & P_{0} {\tilde{r}}^{q/2 - 13/4} \frac{2 {\left( 2 p {\mathcal{A}}^{1/2} \right)}^{p}}{\Gamma {\left( p \right)}} {K_{p} {\left( 4 p {\mathcal{A}}^{1/2} {\tilde{r}}^{- 1 / {\left( 4 p \right)}} \right)}},
\end{eqnarray}
}where $P_{0}$ is a constant.
For the case of $q = 1/2$ and $p = 1$, we obtain the following equation:
\begin{equation}
P = P_{0} {\tilde{r}}^{- 3} \cdot 4 {\mathcal{A}}^{1/2} {K_{1} {\left( 4 {\mathcal{A}}^{1/2} {\tilde{r}}^{- 1/4} \right)}}.
\label{eq:steadyP}
\end{equation}
In this case, $P$ takes the maximum at
\begin{equation}
r = 0.015 {\mathcal{A}}^{2}\ {\rm au}.
%0.014791282091377083189
\end{equation}

Assuming $\mathcal{A} = 26.1$ (see Equation \ref{eq:App2}), the location of the pressure maximum in the steady-state disk is estimated to be $r = 10.1\ {\rm au}$. %10.077322110786790689256504770584405337021
Our numerical simulation shows good agreement with this analytical prediction; the location of the pressure maximum is around $r \simeq 12\ {\rm au}$ at $t = 2\ {\rm Myr}$ (see Figure \ref{fig:gas}).
Thus the structure of the gas disk would already approach the steady-state solution at $t = 2\ {\rm Myr}$.
We note, however, that the location of the maximum for the gas density is not consistent with the analytical prediction.
This is because an exponential cutoff for the outer edge of the gas disk exists at $r \simeq 100\ {\rm au}$ in our numerical simulation (see Equation \ref{eq:Sigmagini}), which is not taken into account in the analytic model for the steady-state disk.

%Similarly, for the case of $q = 0$ (i.e., $p = 1/2$), the stationary solution of $\Sigma_{\rm gas}$ is given by
%\begin{equation}
%\Sigma_{\rm gas} = \Sigma_{0} {\tilde{r}}^{- 7/4} \cdot 2 {\mathcal{A}}^{1/4} {K_{1/2} {\left( 2 {\mathcal{A}}^{1/2} {\tilde{r}}^{- 1/2} \right)}}.
%\end{equation}

%\begin{equation}
%r = \frac{4}{9} \mathcal{A}\ {\rm au}.
%\end{equation}

%\begin{equation}
%P = \frac{\Sigma_{\rm gas} c_{\rm s} \Omega_{\rm K}}{\sqrt{2 \pi}} = p_{0} {\tilde{r}}^{- 13/4} {K_{1/2} {\left( 2 {\mathcal{A}}^{1/2} {\tilde{r}}^{- 1/2} \right)}}.
%\end{equation}

%\begin{equation}
%r = \frac{1}{9} \mathcal{A}\ {\rm au}.
%\end{equation}

\bibliography{sample63}{}

\begin{thebibliography}{}
\expandafter\ifx\csname natexlab\endcsname\relax\def\natexlab#1{#1}\fi
\providecommand{\url}[1]{\href{#1}{#1}}
\providecommand{\dodoi}[1]{doi:~\href{http://doi.org/#1}{\nolinkurl{#1}}}
\providecommand{\doeprint}[1]{\href{http://ascl.net/#1}{\nolinkurl{http://ascl.net/#1}}}
\providecommand{\doarXiv}[1]{\href{https://arxiv.org/abs/#1}{\nolinkurl{https://arxiv.org/abs/#1}}}

\bibitem[{{Adachi} {et~al.}(1976){Adachi}, {Hayashi}, \&
  {Nakazawa}}]{1976PThPh..56.1756A}
{Adachi}, I., {Hayashi}, C., \& {Nakazawa}, K. 1976, Progress of Theoretical
  Physics, 56, 1756, \dodoi{10.1143/PTP.56.1756}

\bibitem[{{ALMA Partnership} {et~al.}(2015){ALMA Partnership}, {Brogan},
  {P{\'e}rez}, {Hunter}, {Dent}, {Hales}, {Hills}, {Corder}, {Fomalont},
  {Vlahakis}, {Asaki}, {Barkats}, {Hirota}, {Hodge}, {Impellizzeri}, {Kneissl},
  {Liuzzo}, {Lucas}, {Marcelino}, {Matsushita}, {Nakanishi}, {Phillips},
  {Richards}, {Toledo}, {Aladro}, {Broguiere}, {Cortes}, {Cortes}, {Espada},
  {Galarza}, {Garcia-Appadoo}, {Guzman-Ramirez}, {Humphreys}, {Jung}, {Kameno},
  {Laing}, {Leon}, {Marconi}, {Mignano}, {Nikolic}, {Nyman}, {Radiszcz},
  {Remijan}, {Rod{\'o}n}, {Sawada}, {Takahashi}, {Tilanus}, {Vila Vilaro},
  {Watson}, {Wiklind}, {Akiyama}, {Chapillon}, {de Gregorio-Monsalvo}, {Di
  Francesco}, {Gueth}, {Kawamura}, {Lee}, {Nguyen Luong}, {Mangum}, {Pietu},
  {Sanhueza}, {Saigo}, {Takakuwa}, {Ubach}, {van Kempen}, {Wootten},
  {Castro-Carrizo}, {Francke}, {Gallardo}, {Garcia}, {Gonzalez}, {Hill},
  {Kaminski}, {Kurono}, {Liu}, {Lopez}, {Morales}, {Plarre}, {Schieven},
  {Testi}, {Videla}, {Villard}, {Andreani}, {Hibbard}, \&
  {Tatematsu}}]{2015ApJ...808L...3A}
{ALMA Partnership}, {Brogan}, C.~L., {P{\'e}rez}, L.~M., {et~al.} 2015, \apjl,
  808, L3, \dodoi{10.1088/2041-8205/808/1/L3}

\bibitem[{{Andrews}(2020)}]{2020ARA&A..58..483A}
{Andrews}, S.~M. 2020, \araa, 58, 483,
  \dodoi{10.1146/annurev-astro-031220-010302}

\bibitem[{{Andrews} {et~al.}(2018){Andrews}, {Huang}, {P{\'e}rez}, {Isella},
  {Dullemond}, {Kurtovic}, {Guzm{\'a}n}, {Carpenter}, {Wilner}, {Zhang}, {Zhu},
  {Birnstiel}, {Bai}, {Benisty}, {Hughes}, {{\"O}berg}, \&
  {Ricci}}]{2018ApJ...869L..41A}
{Andrews}, S.~M., {Huang}, J., {P{\'e}rez}, L.~M., {et~al.} 2018, \apjl, 869,
  L41, \dodoi{10.3847/2041-8213/aaf741}

\bibitem[{{Bai} \& {Stone}(2010{\natexlab{a}})}]{2010ApJ...722L.220B}
{Bai}, X.-N., \& {Stone}, J.~M. 2010{\natexlab{a}}, \apjl, 722, L220,
  \dodoi{10.1088/2041-8205/722/2/L220}

\bibitem[{{Bai} \& {Stone}(2010{\natexlab{b}})}]{2010ApJ...722.1437B}
---. 2010{\natexlab{b}}, \apj, 722, 1437, \dodoi{10.1088/0004-637X/722/2/1437}

\bibitem[{{Boss}(2008)}]{2008E&PSL.268..102B}
{Boss}, A.~P. 2008, Earth and Planetary Science Letters, 268, 102,
  \dodoi{10.1016/j.epsl.2008.01.008}

\bibitem[{{Carballido} {et~al.}(2005){Carballido}, {Stone}, \&
  {Pringle}}]{2005MNRAS.358.1055C}
{Carballido}, A., {Stone}, J.~M., \& {Pringle}, J.~E. 2005, \mnras, 358, 1055,
  \dodoi{10.1111/j.1365-2966.2005.08850.x}

\bibitem[{{Carrera} {et~al.}(2015){Carrera}, {Johansen}, \&
  {Davies}}]{2015A&A...579A..43C}
{Carrera}, D., {Johansen}, A., \& {Davies}, M.~B. 2015, \aap, 579, A43,
  \dodoi{10.1051/0004-6361/201425120}

\bibitem[{{Ciesla}(2007)}]{2007Sci...318..613C}
{Ciesla}, F.~J. 2007, Science, 318, 613, \dodoi{10.1126/science.1147273}

\bibitem[{{Ciesla}(2009)}]{2009Icar..200..655C}
---. 2009, \icarus, 200, 655, \dodoi{10.1016/j.icarus.2008.12.009}

\bibitem[{{Ciesla}(2010)}]{2010Icar..208..455C}
---. 2010, \icarus, 208, 455, \dodoi{10.1016/j.icarus.2010.02.010}

\bibitem[{{Clarke} \& {Pringle}(1988)}]{1988MNRAS.235..365C}
{Clarke}, C.~J., \& {Pringle}, J.~E. 1988, \mnras, 235, 365,
  \dodoi{10.1093/mnras/235.2.365}

\bibitem[{{de Vries} {et~al.}(2012){de Vries}, {Acke}, {Blommaert}, {Waelkens},
  {Waters}, {Vandenbussche}, {Min}, {Olofsson}, {Dominik}, {Decin}, {Barlow},
  {Brandeker}, {di Francesco}, {Glauser}, {Greaves}, {Harvey}, {Holland},
  {Ivison}, {Liseau}, {Pantin}, {Pilbratt}, {Royer}, \&
  {Sibthorpe}}]{2012Natur.490...74D}
{de Vries}, B.~L., {Acke}, B., {Blommaert}, J.~A.~D.~L., {et~al.} 2012, \nat,
  490, 74, \dodoi{10.1038/nature11469}

\bibitem[{{Desch} {et~al.}(2017){Desch}, {Estrada}, {Kalyaan}, \&
  {Cuzzi}}]{2017ApJ...840...86D}
{Desch}, S.~J., {Estrada}, P.~R., {Kalyaan}, A., \& {Cuzzi}, J.~N. 2017, \apj,
  840, 86, \dodoi{10.3847/1538-4357/aa6bfb}

\bibitem[{{Desch} {et~al.}(2018){Desch}, {Kalyaan}, \& {O'D.
  Alexander}}]{2018ApJS..238...11D}
{Desch}, S.~J., {Kalyaan}, A., \& {O'D. Alexander}, C.~M. 2018, \apjs, 238, 11,
  \dodoi{10.3847/1538-4365/aad95f}

\bibitem[{{Dong} {et~al.}(2015){Dong}, {Zhu}, \&
  {Whitney}}]{2015ApJ...809...93D}
{Dong}, R., {Zhu}, Z., \& {Whitney}, B. 2015, \apj, 809, 93,
  \dodoi{10.1088/0004-637X/809/1/93}

\bibitem[{{Dr{\k{a}}{\.z}kowska} {et~al.}(2016){Dr{\k{a}}{\.z}kowska},
  {Alibert}, \& {Moore}}]{2016A&A...594A.105D}
{Dr{\k{a}}{\.z}kowska}, J., {Alibert}, Y., \& {Moore}, B. 2016, \aap, 594,
  A105, \dodoi{10.1051/0004-6361/201628983}

\bibitem[{{Dullemond} {et~al.}(2006){Dullemond}, {Apai}, \&
  {Walch}}]{2006ApJ...640L..67D}
{Dullemond}, C.~P., {Apai}, D., \& {Walch}, S. 2006, \apjl, 640, L67,
  \dodoi{10.1086/503100}

\bibitem[{{Ercolano} \& {Pascucci}(2017)}]{2017RSOS....470114E}
{Ercolano}, B., \& {Pascucci}, I. 2017, Royal Society Open Science, 4, 170114,
  \dodoi{10.1098/rsos.170114}

\bibitem[{{Fukagawa} {et~al.}(2013){Fukagawa}, {Tsukagoshi}, {Momose}, {Saigo},
  {Ohashi}, {Kitamura}, {Inutsuka}, {Muto}, {Nomura}, {Takeuchi}, {Kobayashi},
  {Hanawa}, {Akiyama}, {Honda}, {Fujiwara}, {Kataoka}, {Takahashi}, \&
  {Shibai}}]{2013PASJ...65L..14F}
{Fukagawa}, M., {Tsukagoshi}, T., {Momose}, M., {et~al.} 2013, \pasj, 65, L14,
  \dodoi{10.1093/pasj/65.6.L14}

\bibitem[{{Fukai} \& {Arakawa}(2021)}]{2021ApJ...908...64F}
{Fukai}, R., \& {Arakawa}, S. 2021, \apj, 908, 64,
  \dodoi{10.3847/1538-4357/abd2b9}

\bibitem[{{Gail}(2001)}]{2001A&A...378..192G}
{Gail}, H.~P. 2001, \aap, 378, 192, \dodoi{10.1051/0004-6361:20011130}

\bibitem[{{Haghighipour} \& {Boss}(2003)}]{2003ApJ...583..996H}
{Haghighipour}, N., \& {Boss}, A.~P. 2003, \apj, 583, 996,
  \dodoi{10.1086/345472}

\bibitem[{{Hallenbeck} {et~al.}(2000){Hallenbeck}, {Nuth}, \&
  {Nelson}}]{2000ApJ...535..247H}
{Hallenbeck}, S.~L., {Nuth}, Joseph~A., I., \& {Nelson}, R.~N. 2000, \apj, 535,
  247, \dodoi{10.1086/308810}

\bibitem[{{Hanner} {et~al.}(1995){Hanner}, {Brooke}, \&
  {Tokunaga}}]{1995ApJ...438..250H}
{Hanner}, M.~S., {Brooke}, T.~Y., \& {Tokunaga}, A.~T. 1995, \apj, 438, 250,
  \dodoi{10.1086/175069}

\bibitem[{{Harker} \& {Desch}(2002)}]{2002ApJ...565L.109H}
{Harker}, D.~E., \& {Desch}, S.~J. 2002, \apjl, 565, L109,
  \dodoi{10.1086/339363}

\bibitem[{{Hartmann} {et~al.}(1998){Hartmann}, {Calvet}, {Gullbring}, \&
  {D'Alessio}}]{1998ApJ...495..385H}
{Hartmann}, L., {Calvet}, N., {Gullbring}, E., \& {D'Alessio}, P. 1998, \apj,
  495, 385, \dodoi{10.1086/305277}

\bibitem[{{Honda} {et~al.}(2003){Honda}, {Kataza}, {Okamoto}, {Miyata},
  {Yamashita}, {Sako}, {Takubo}, \& {Onaka}}]{2003ApJ...585L..59H}
{Honda}, M., {Kataza}, H., {Okamoto}, Y.~K., {et~al.} 2003, \apjl, 585, L59,
  \dodoi{10.1086/374034}

\bibitem[{{Honda} {et~al.}(2004){Honda}, {Watanabe}, {Yamashita}, {Kataza},
  {Okamoto}, {Miyata}, {Sako}, {Fujiyoshi}, {Kawakita}, {Furusho}, {Kinoshita},
  {Sekiguchi}, {Ootsubo}, \& {Onaka}}]{2004ApJ...601..577H}
{Honda}, M., {Watanabe}, J.-i., {Yamashita}, T., {et~al.} 2004, \apj, 601, 577,
  \dodoi{10.1086/380478}

\bibitem[{{Honda} {et~al.}(2006){Honda}, {Kataza}, {Okamoto}, {Yamashita},
  {Min}, {Miyata}, {Sako}, {Fujiyoshi}, {Sakon}, \&
  {Onaka}}]{2006ApJ...646.1024H}
{Honda}, M., {Kataza}, H., {Okamoto}, Y.~K., {et~al.} 2006, \apj, 646, 1024,
  \dodoi{10.1086/505035}

\bibitem[{{Hughes} \& {Armitage}(2010)}]{2010ApJ...719.1633H}
{Hughes}, A. L.~H., \& {Armitage}, P.~J. 2010, \apj, 719, 1633,
  \dodoi{10.1088/0004-637X/719/2/1633}

\bibitem[{{Johansen} {et~al.}(2014){Johansen}, {Blum}, {Tanaka}, {Ormel},
  {Bizzarro}, \& {Rickman}}]{2014prpl.conf..547J}
{Johansen}, A., {Blum}, J., {Tanaka}, H., {et~al.} 2014, in Protostars and
  Planets VI, ed. H.~{Beuther}, R.~S. {Klessen}, C.~P. {Dullemond}, \&
  T.~{Henning}, 547, \dodoi{10.2458/azu\_uapress\_9780816531240-ch024}

\bibitem[{{Johansen} \& {Klahr}(2005)}]{2005ApJ...634.1353J}
{Johansen}, A., \& {Klahr}, H. 2005, \apj, 634, 1353, \dodoi{10.1086/497118}

\bibitem[{{Juh{\'a}sz} {et~al.}(2010){Juh{\'a}sz}, {Bouwman}, {Henning},
  {Acke}, {van den Ancker}, {Meeus}, {Dominik}, {Min}, {Tielens}, \&
  {Waters}}]{2010ApJ...721..431J}
{Juh{\'a}sz}, A., {Bouwman}, J., {Henning}, T., {et~al.} 2010, \apj, 721, 431,
  \dodoi{10.1088/0004-637X/721/1/431}

\bibitem[{{Kanagawa} {et~al.}(2018){Kanagawa}, {Muto}, {Okuzumi}, {Tanigawa},
  {Taki}, \& {Shibaike}}]{2018ApJ...868...48K}
{Kanagawa}, K.~D., {Muto}, T., {Okuzumi}, S., {et~al.} 2018, \apj, 868, 48,
  \dodoi{10.3847/1538-4357/aae837}

\bibitem[{{Keller} \& {Gail}(2004)}]{2004A&A...415.1177K}
{Keller}, C., \& {Gail}, H.~P. 2004, \aap, 415, 1177,
  \dodoi{10.1051/0004-6361:20034629}

\bibitem[{{Kemper} {et~al.}(2004){Kemper}, {Vriend}, \&
  {Tielens}}]{2004ApJ...609..826K}
{Kemper}, F., {Vriend}, W.~J., \& {Tielens}, A.~G.~G.~M. 2004, \apj, 609, 826,
  \dodoi{10.1086/421339}

\bibitem[{{Lambrechts} \& {Johansen}(2014)}]{2014A&A...572A.107L}
{Lambrechts}, M., \& {Johansen}, A. 2014, \aap, 572, A107,
  \dodoi{10.1051/0004-6361/201424343}

\bibitem[{{Lynden-Bell} \& {Pringle}(1974)}]{1974MNRAS.168..603L}
{Lynden-Bell}, D., \& {Pringle}, J.~E. 1974, \mnras, 168, 603,
  \dodoi{10.1093/mnras/168.3.603}

\bibitem[{{Maaskant} {et~al.}(2015){Maaskant}, {de Vries}, {Min}, {Waters},
  {Dominik}, {Molster}, \& {Tielens}}]{2015A&A...574A.140M}
{Maaskant}, K.~M., {de Vries}, B.~L., {Min}, M., {et~al.} 2015, \aap, 574,
  A140, \dodoi{10.1051/0004-6361/201423770}

\bibitem[{{Miura} {et~al.}(2010){Miura}, {Tanaka}, {Yamamoto}, {Nakamoto},
  {Yamada}, {Tsukamoto}, \& {Nozawa}}]{2010ApJ...719..642M}
{Miura}, H., {Tanaka}, K.~K., {Yamamoto}, T., {et~al.} 2010, \apj, 719, 642,
  \dodoi{10.1088/0004-637X/719/1/642}

\bibitem[{{Miyake} {et~al.}(2016){Miyake}, {Suzuki}, \&
  {Inutsuka}}]{2016ApJ...821....3M}
{Miyake}, T., {Suzuki}, T.~K., \& {Inutsuka}, S.-i. 2016, \apj, 821, 3,
  \dodoi{10.3847/0004-637X/821/1/3}

\bibitem[{{Mousis} {et~al.}(2007){Mousis}, {Petit}, {Wurm}, {Krauss},
  {Alibert}, \& {Horner}}]{2007A&A...466L...9M}
{Mousis}, O., {Petit}, J.~M., {Wurm}, G., {et~al.} 2007, \aap, 466, L9,
  \dodoi{10.1051/0004-6361:20077170}

\bibitem[{{Ogliore} {et~al.}(2009){Ogliore}, {Westphal}, {Gainsforth},
  {Butterworth}, {Fakra}, \& {Marcus}}]{2009M&PS...44.1675O}
{Ogliore}, R.~C., {Westphal}, A.~J., {Gainsforth}, Z., {et~al.} 2009,
  Meteoritics and Planetary Science, 44, 1675,
  \dodoi{10.1111/j.1945-5100.2009.tb01198.x}

\bibitem[{{Ohashi} {et~al.}(2021){Ohashi}, {Kobayashi}, {Nakatani}, {Okuzumi},
  {Tanaka}, {Murakawa}, {Zhang}, {Liu}, \& {Sakai}}]{2021ApJ...907...80O}
{Ohashi}, S., {Kobayashi}, H., {Nakatani}, R., {et~al.} 2021, \apj, 907, 80,
  \dodoi{10.3847/1538-4357/abd0fa}

\bibitem[{{Okuzumi} {et~al.}(2016){Okuzumi}, {Momose}, {Sirono}, {Kobayashi},
  \& {Tanaka}}]{2016ApJ...821...82O}
{Okuzumi}, S., {Momose}, M., {Sirono}, S.-i., {Kobayashi}, H., \& {Tanaka}, H.
  2016, \apj, 821, 82, \dodoi{10.3847/0004-637X/821/2/82}

\bibitem[{{Okuzumi} {et~al.}(2012){Okuzumi}, {Tanaka}, {Kobayashi}, \&
  {Wada}}]{2012ApJ...752..106O}
{Okuzumi}, S., {Tanaka}, H., {Kobayashi}, H., \& {Wada}, K. 2012, \apj, 752,
  106, \dodoi{10.1088/0004-637X/752/2/106}

\bibitem[{{Okuzumi} \& {Tazaki}(2019)}]{2019ApJ...878..132O}
{Okuzumi}, S., \& {Tazaki}, R. 2019, \apj, 878, 132,
  \dodoi{10.3847/1538-4357/ab204d}

\bibitem[{{Ootsubo} {et~al.}(2007){Ootsubo}, {Watanabe}, {Kawakita}, {Honda},
  \& {Furusho}}]{2007P&SS...55.1044O}
{Ootsubo}, T., {Watanabe}, J.-i., {Kawakita}, H., {Honda}, M., \& {Furusho}, R.
  2007, \planss, 55, 1044, \dodoi{10.1016/j.pss.2006.11.012}

\bibitem[{{Ormel} \& {Cuzzi}(2007)}]{2007A&A...466..413O}
{Ormel}, C.~W., \& {Cuzzi}, J.~N. 2007, \aap, 466, 413,
  \dodoi{10.1051/0004-6361:20066899}

\bibitem[{{Pavlyuchenkov} \& {Dullemond}(2007)}]{2007A&A...471..833P}
{Pavlyuchenkov}, Y., \& {Dullemond}, C.~P. 2007, \aap, 471, 833,
  \dodoi{10.1051/0004-6361:20077317}

\bibitem[{{Pinilla} {et~al.}(2017){Pinilla}, {Pohl}, {Stammler}, \&
  {Birnstiel}}]{2017ApJ...845...68P}
{Pinilla}, P., {Pohl}, A., {Stammler}, S.~M., \& {Birnstiel}, T. 2017, \apj,
  845, 68, \dodoi{10.3847/1538-4357/aa7edb}

\bibitem[{{Sekiya} \& {Onishi}(2018)}]{2018ApJ...860..140S}
{Sekiya}, M., \& {Onishi}, I.~K. 2018, \apj, 860, 140,
  \dodoi{10.3847/1538-4357/aac4a7}

\bibitem[{{Shakura} \& {Sunyaev}(1973)}]{1973A&A....24..337S}
{Shakura}, N.~I., \& {Sunyaev}, R.~A. 1973, \aap, 500, 33

\bibitem[{{Sturm} {et~al.}(2013){Sturm}, {Bouwman}, {Henning}, {Evans},
  {Waters}, {van Dishoeck}, {Green}, {Olofsson}, {Meeus}, {Maaskant},
  {Dominik}, {Augereau}, {Mulders}, {Acke}, {Merin}, \&
  {Herczeg}}]{2013A&A...553A...5S}
{Sturm}, B., {Bouwman}, J., {Henning}, T., {et~al.} 2013, \aap, 553, A5,
  \dodoi{10.1051/0004-6361/201220243}

\bibitem[{{Suzuki} \& {Inutsuka}(2009)}]{2009ApJ...691L..49S}
{Suzuki}, T.~K., \& {Inutsuka}, S.-i. 2009, \apjl, 691, L49,
  \dodoi{10.1088/0004-637X/691/1/L49}

\bibitem[{{Suzuki} {et~al.}(2010){Suzuki}, {Muto}, \&
  {Inutsuka}}]{2010ApJ...718.1289S}
{Suzuki}, T.~K., {Muto}, T., \& {Inutsuka}, S.-i. 2010, \apj, 718, 1289,
  \dodoi{10.1088/0004-637X/718/2/1289}

\bibitem[{{Suzuki} {et~al.}(2016){Suzuki}, {Ogihara}, {Morbidelli}, {Crida}, \&
  {Guillot}}]{2016A&A...596A..74S}
{Suzuki}, T.~K., {Ogihara}, M., {Morbidelli}, A., {Crida}, A., \& {Guillot}, T.
  2016, \aap, 596, A74, \dodoi{10.1051/0004-6361/201628955}

\bibitem[{{Takahashi} \& {Inutsuka}(2014)}]{2014ApJ...794...55T}
{Takahashi}, S.~Z., \& {Inutsuka}, S.-i. 2014, \apj, 794, 55,
  \dodoi{10.1088/0004-637X/794/1/55}

\bibitem[{{Takahashi} \& {Muto}(2018)}]{2018ApJ...865..102T}
{Takahashi}, S.~Z., \& {Muto}, T. 2018, \apj, 865, 102,
  \dodoi{10.3847/1538-4357/aadda0}

\bibitem[{{Taki} {et~al.}(2021){Taki}, {Kuwabara}, {Kobayashi}, \&
  {Suzuki}}]{2021ApJ...909...75T}
{Taki}, T., {Kuwabara}, K., {Kobayashi}, H., \& {Suzuki}, T.~K. 2021, \apj,
  909, 75, \dodoi{10.3847/1538-4357/abd79f}

\bibitem[{{Tanaka} {et~al.}(2010){Tanaka}, {Yamamoto}, \&
  {Kimura}}]{2010ApJ...717..586T}
{Tanaka}, K.~K., {Yamamoto}, T., \& {Kimura}, H. 2010, \apj, 717, 586,
  \dodoi{10.1088/0004-637X/717/1/586}

\bibitem[{{Tazaki} \& {Nomura}(2015)}]{2015ApJ...799..119T}
{Tazaki}, R., \& {Nomura}, H. 2015, \apj, 799, 119,
  \dodoi{10.1088/0004-637X/799/2/119}

\bibitem[{{Testi} {et~al.}(2014){Testi}, {Birnstiel}, {Ricci}, {Andrews},
  {Blum}, {Carpenter}, {Dominik}, {Isella}, {Natta}, {Williams}, \&
  {Wilner}}]{2014prpl.conf..339T}
{Testi}, L., {Birnstiel}, T., {Ricci}, L., {et~al.} 2014, in Protostars and
  Planets VI, ed. H.~{Beuther}, R.~S. {Klessen}, C.~P. {Dullemond}, \&
  T.~{Henning}, 339, \dodoi{10.2458/azu\_uapress\_9780816531240-ch015}

\bibitem[{{Tominaga} {et~al.}(2019){Tominaga}, {Takahashi}, \&
  {Inutsuka}}]{2019ApJ...881...53T}
{Tominaga}, R.~T., {Takahashi}, S.~Z., \& {Inutsuka}, S.-i. 2019, \apj, 881,
  53, \dodoi{10.3847/1538-4357/ab25ea}

\bibitem[{{Tsukagoshi} {et~al.}(2016){Tsukagoshi}, {Nomura}, {Muto}, {Kawabe},
  {Ishimoto}, {Kanagawa}, {Okuzumi}, {Ida}, {Walsh}, \&
  {Millar}}]{2016ApJ...829L..35T}
{Tsukagoshi}, T., {Nomura}, H., {Muto}, T., {et~al.} 2016, \apjl, 829, L35,
  \dodoi{10.3847/2041-8205/829/2/L35}

\bibitem[{{Ueda} {et~al.}(2019){Ueda}, {Flock}, \&
  {Okuzumi}}]{2019ApJ...871...10U}
{Ueda}, T., {Flock}, M., \& {Okuzumi}, S. 2019, \apj, 871, 10,
  \dodoi{10.3847/1538-4357/aaf3a1}

\bibitem[{{van Boekel} {et~al.}(2017){van Boekel}, {Henning}, {Menu}, {de
  Boer}, {Langlois}, {M{\"u}ller}, {Avenhaus}, {Boccaletti}, {Schmid},
  {Thalmann}, {Benisty}, {Dominik}, {Ginski}, {Girard}, {Gisler}, {Lobo Gomes},
  {Menard}, {Min}, {Pavlov}, {Pohl}, {Quanz}, {Rabou}, {Roelfsema}, {Sauvage},
  {Teague}, {Wildi}, \& {Zurlo}}]{2017ApJ...837..132V}
{van Boekel}, R., {Henning}, T., {Menu}, J., {et~al.} 2017, \apj, 837, 132,
  \dodoi{10.3847/1538-4357/aa5d68}

\bibitem[{{Vinkovi{\'c}}(2009)}]{2009Natur.459..227V}
{Vinkovi{\'c}}, D. 2009, \nat, 459, 227, \dodoi{10.1038/nature08032}

\bibitem[{{Yamamoto} \& {Tachibana}(2018)}]{2018ECS.....2..778Y}
{Yamamoto}, D., \& {Tachibana}, S. 2018, ACS Earth and Space Chemistry, 2, 778,
  \dodoi{10.1021/acsearthspacechem.8b00047}

\bibitem[{{Yang} {et~al.}(2017){Yang}, {Johansen}, \&
  {Carrera}}]{2017A&A...606A..80Y}
{Yang}, C.~C., {Johansen}, A., \& {Carrera}, D. 2017, \aap, 606, A80,
  \dodoi{10.1051/0004-6361/201630106}

\bibitem[{{Yang} \& {Ciesla}(2012)}]{2012M&PS...47...99Y}
{Yang}, L., \& {Ciesla}, F.~J. 2012, Meteoritics and Planetary Science, 47, 99,
  \dodoi{10.1111/j.1945-5100.2011.01315.x}

\bibitem[{{Youdin} \& {Lithwick}(2007)}]{2007Icar..192..588Y}
{Youdin}, A.~N., \& {Lithwick}, Y. 2007, \icarus, 192, 588,
  \dodoi{10.1016/j.icarus.2007.07.012}

\end{thebibliography}
\bibliographystyle{aasjournal}

%% This command is needed to show the entire author+affiliation list when
%% the collaboration and author truncation commands are used.  It has to
%% go at the end of the manuscript.
%\allauthors

%% Include this line if you are using the \added, \replaced, \deleted
%% commands to see a summary list of all changes at the end of the article.
%\listofchanges

\listofchanges
\end{document}